\documentclass[12pt]{article}
 \usepackage{epsfig}
\usepackage[latin1]{inputenc}

\def\lsim{\,\raise0.3ex\hbox{$<$\kern-0.75em\raise-1.1ex\hbox{$\sim$}}\,}
\def\gsim{\,\raise0.3ex\hbox{$>$\kern-0.75em\raise-1.1ex\hbox{$\sim$}}\,}

\def\Teff{T_{\rm eff}}
\def\psat{p_{\rm sat}}

\def\ssNN{\sqrt{s_{NN}}}

\def\Teff{T_{\rm eff}}

\def\Tdec{T_{\rm dec}}

\newcommand{\beq}{\begin{equation}}
\newcommand{\eeq}{\end{equation}}
\newcommand{\bea}{\begin{eqnarray}}
\newcommand{\eea}{\end{eqnarray}}

\begin{document}

\begin{titlepage}
\begin{flushright}
%Draft 22\\
2 June, 2005\\
HIP-2005-24/TH\\
hep-ph/0506049
\end{flushright}
\begin{centering}
\vfill

{\large \bf RHIC-tested predictions for low-$p_T$ and high-$p_T$ hadron spectra
in nearly central Pb+Pb collisions at the LHC}

\vspace{0.5cm}
 K. J. Eskola,$^{\rm a,b,}$\footnote{
kari.eskola, heli.honkanen, harri.niemi, vesa.ruuskanen,
sami.rasanen@phys.jyu.fi
}
 H. Honkanen,$^{\rm a,b,c,1}$
 H. Niemi,$^{\rm a,b,1}$\\
 P. V. Ruuskanen,$^{\rm a,b,1}$ and
 S. S. R\"as\"anen$^{\rm a,1}$

\vspace{1cm}
{\em $^{\rm a}$Department of Physics,
P.B. 35, FIN-40014 University of Jyv\"askyl\"a, Finland}\\
\vspace{0.3cm}
{\em $^{\rm b}$Helsinki Institute of Physics,
P.B. 64, FIN-00014 University of Helsinki, Finland}\\
\vspace{0.3cm}
{\em $^{\rm c}$Department of Physics, University of Virginia,
P.O.B. 400714, Charlottesville, VA 22904-4714, USA}\\

\vspace{1cm}\centerline{\bf Abstract}
\end{centering}

%%%%%% ABSTRACT:
We study the hadron spectra in nearly central $A$+$A$ collisions
at RHIC and LHC in a broad transverse momentum range.
%%% low-PT
We cover the low-$p_T$ spectra using longitudinally boost-invariant 
hydrodynamics with initial energy and net-baryon number densities from 
the perturbative QCD (pQCD)+saturation model. Build-up of the transverse
flow and sensitivity of the spectra to a single decoupling temperature
$\Tdec$ are studied. Comparison with RHIC data at $\ssNN=130$ and 200 
GeV  suggests a rather high value $\Tdec=150$~MeV.
%%% high-pT
The high-$p_T$ spectra are computed using factorized pQCD cross 
sections, nuclear parton distributions, fragmentation functions, and 
describing partonic energy loss in the quark-gluon plasma by quenching 
weights. Overall normalization is fixed on the basis of p+$\bar{\rm p}$(p) 
data and the strength of energy loss is determined from RHIC Au+Au data. 
Uncertainties are discussed.
%%% LHC
With constraints from RHIC data, we predict the $p_T$ spectra of hadrons
in 5 \% most central Pb+Pb collisions at the LHC energy 
$\ssNN=5500$~GeV. Due to the closed framework for primary production, we 
can also predict the net-baryon number at midrapidity, as well as the 
strength of partonic energy losses at the LHC.
%%% hydro vs. pQCD fragmentation.
Both at the LHC and RHIC, we recognize a rather narrow crossover region
in the $p_T$ spectra, where the hydrodynamic and pQCD fragmentation
components become of equal size. We argue that in this crossover region 
the two contributions are to a good approximation mutually independent.
In particular, our results suggest a wider $p_T$-region of applicability
for hydrodynamical models at the LHC than at RHIC.

\vfill
\end{titlepage}

%%%%%%%%%%%%%%%%%%%%%%%%%%%%%%%
%%%%  INTRODUCTION
%%%%%%%%%%%%%%%%%%%%%%%%%%%%%%%
\section{Introduction}

Transverse momentum spectra of hadrons in ultrarelativistic heavy  ion
collisions offer invaluable input for the studies of QCD matter
properties and QCD dynamics. At high collision energies, currently at
BNL-RHIC and in the near future at the CERN-LHC, the parton and hadron
production rates are significant over a wide transverse momentum
range. The high- and low-$p_T$ parts of the transverse
momentum spectra provide complementary information on the QCD
matter. As reviewed below, the hadron spectra measured in Au+Au
collisions at RHIC show systematic evidence on formation of
elementary-particle matter and provide motivation for hydrodynamical
description of $A$+$A$ collisions.

\paragraph{Large energy densities.}
First, average initial energy densities can be estimated on the
basis of measured transverse energy \cite{ETmeas} originating from
the low-$p_T$ region of the hadron spectra.  Even the conservative
Bjorken estimate \cite{BJORKEN}, $\epsilon=\frac{dE_T}{d\eta}/(\pi
R_A^2\tau)$, computed using $dE_T/d\eta= 606\dots620$~GeV as
measured at $\eta\sim0$ in $0\dots5$ \% central Au+Au collisions at
$\ssNN=200$~GeV at RHIC, implies an average local energy density
$\epsilon\approx 5$~GeV/fm$^{3}$ \cite{ETmeas} at $\tau\sim 1$~fm, a
typical hadronization time in a p+p collision in vacuum.  Already
this is well above the energy densities where hadrons could
form. Since the formation time of such a system at RHIC is most likely
considerably shorter than 1 fm, the average initial energy densities
can be expected to be clearly, say, even a factor five, higher.
Furthermore, if there is pressure in the system, the estimates of
energy density become even larger, since $dE_T/dy$ is not conserved
due to the $pdV$ work by pressure in longitudinal
expansion. According to the hydrodynamic studies, the initial-to-final
transverse energy degradation can be as much as a factor three
\cite{ERRT}, increasing the initial energy density estimates accordingly.

\paragraph{Pressure.}
Second, clear azimuthal asymmetry in the transverse spectra has been
observed \cite{ellipticflow} in non-central collisions. This
phenomenon is possible only if sufficiently frequent secondary
collisions take place to build up pressure. Furthermore, hydrodynamic
studies imply that the pressure must form early enough (in less than 1~fm) for
the spatial azimuthal asymmetry in the primary production to be
transferred into momentum distributions \cite{KOLB}. Complete
thermalization, however, is not necessary for generating pressure in
the system. Both observations, the large energy densities and
the early existence of pressure, considered together suggest that
a collectively behaving partonic matter of gluons, quarks and
antiquarks, is produced.

\paragraph{Partonic energy loss.}
Third, the high-$p_T$ hadron spectra in Au+Au collisions at $p_T\gsim
4\dots 5$ GeV have been  observed to be suppressed  by a factor $\sim 4\dots 5$
relative to p+p \cite{Adams:2003kv,Adler:2002xw,Adler:2003au,Adcox:2001jp,
Back:2003qr,Arsene:2003yk}. This energy loss, jet quenching, is one of
the predicted QGP signals \cite{JETPROCS}. The fact that no such
suppression is discovered in d+Au collisions \cite{dAu}, excludes
the possibility that the primary QCD collision mechanism for producing
high-$p_T$ partons in nuclear collisions would be very different from
that in p+p. Also the back-to-back correlation of the high-$p_T$
hadrons, present in the p+p collisions, has been observed to disappear
in central Au+Au \cite{AWAYSIDE}. These results, considered together
with the first two, very strongly suggest that the origin of the
observed high-$p_T$ suppressions in Au+Au is the produced partonic
medium: before hadronization the hard partons lose energy in
penetrating through the dense parton matter.

\paragraph{Hydro for low-$p_T$ and pQCD for high-$p_T$ spectra.}
The interpretation of the HBT radii extracted from the particle
correlation measurements \cite{HBTpuzzle} is still a puzzle to be
solved for a better understanding of the produced collective
system. In spite of this caveat, the RHIC results indicate that a
dense strongly interacting partonic system, where mean free paths are
much smaller than the size of the system, is produced in nonperipheral
$A$+$A$ collisions. It is then reasonable to expect that hadron spectra
at $p_T\lsim 2...3$~GeV, the region which dominates the total
multiplicity and transverse energy, can be obtained by treating the
expansion of produced matter quantitatively in terms of relativistic
hydrodynamics \cite{KATAJA}. Several pieces of input are necessary for
such calculation: initial state of the evolution must be given in
detail, an Equation of State (EoS) is needed to close the system of
hydrodynamic equations and finally the collective description of
flowing matter must be turned into particle spectra.

In the hydrodynamic approach also the originally produced high-$p_T$
partons are assumed to be absorbed into the thermalized system and all
final state hadrons are thermally emitted from the decoupling
surface. Especially, no remnants of pQCD power-law tails are left
on top of the steeply falling thermal spectra of hadrons.
In the spectra of hadrons measured in Au+Au collisions at RHIC, the
powerlaw like tails do appear at large transverse momenta, $p_T\gsim
5..6$~GeV, but since the high-$p_T$ partons carry only a small
fraction of the total (transverse) energy, the assumption that all partons 
thermalize does not affect the treatment of the thermal part appreciably. 
For the high-$p_T$ region, a more relevant baseline calculation is hadron
production without rescatterings, where all the produced high-$p_T$
partons cascade in vacuum into non-interacting partonic showers which
then independently hadronize. Such reference spectra can be computed
for $A$+$A$ collisions using perturbative QCD (pQCD) in the same way as
in p+$\bar{\rm p}$ and p+p collisions. Energy losses can also be
incorporated into these baseline spectra \cite{JETPROCS}.

\paragraph{In this paper}
we aim at a comprehensive description of hadron spectra in central
$A$+$A$ collisions, comparing the results with RHIC data and giving
detailed predictions for the 5 \% most central Pb+Pb collisions at the
LHC. On one hand, we compute the spectra within the framework of
hydrodynamics, applying initial conditions predicted by pQCD and
saturation \cite{EKRT,ERRT}, and also including decays of
produced resonances after decoupling. On the other hand, we compute
the high-$p_T$ reference spectra in the framework of collinearly
factorized pQCD cross sections, nuclear parton distributions (nPDFs) and
fragmentation functions \cite{EH03}. Following Ref. \cite{EHSW04},
energy losses of high-$p_T$ partons are incorporated in terms of
quenching weights \cite{SW03}. A comparison of the computed low- and
high-$p_T$ spectra with RHIC data is presented and, in
particular, these spectra are predicted for the 5\% most central Pb+Pb
collisions at the LHC. We also determine and discuss the theoretically
challenging crossover region where the hydrodynamic and the pQCD
spectra are comparable.

The rest of this paper is arranged as follows: In
Sec.~\ref{sec:inistate}, we first briefly review the saturation models
from which the initial state for hydrodynamics can be computed at
collider energies. After this, we discuss in more detail the
pQCD+saturation initial conditions. In Sec.~\ref{sec:hydro} we define
the hydrodynamic framework. In particular, we discuss the fast
thermalization assumption, the EoS, and collective flow
phenomena. Section~\ref{sec:frag} specifies the pQCD computation of
the baseline high-$p_T$ spectra without and with energy losses. The
results are presented in Sec.~\ref{sec:results}, and conclusions in
Sec.~\ref{sec:conclusions}.

%%%%%%%%%%%%%%%%%%%%%%%%%%%%%%%%%%
%%%   INITIAL STATE
%%%%%%%%%%%%%%%%%%%%%%%%%%%%%%%%%%
\section{Initial state for hydrodynamics}\label{sec:inistate}
A key input for hydrodynamical studies is the initial state of
evolution. At the SPS energies, hadron multiplicities and spectra have
been used to constrain the initial state, the EoS, flow and decoupling
\cite{SPShydro,Sollfrank_prc}. However, it turned out that different
combinations of initial states and EoS can lead to similar results for
the final spectra. Hence, the spectra could obviously be more
efficiently used to extract information on the properties of QCD
matter and its spacetime evolution if the initial state could be
extracted from elsewhere. This can be expected to be the case at
collider energies as perturbatively computable parton production may
dominate initial parton production in $A$+$A$ collisions at sufficiently
high collider energies \cite{BM87,KLL,EKL89,HIJING}. Due to the
diverging partonic cross sections pQCD alone, however, does not
predict the amount of transverse energy deposited into the central
rapidity region. A further element, like gluon saturation, is
needed. We briefly comment on the saturation models which provide a
basis for estimating the initial conditions.

\subsection{A brief review of saturation models}

\subsubsection{The general idea in saturation - evidence in DIS}

Gluon saturation was introduced more than two decades ago for p+p
collisions \cite{GLR} as a property of the wave functions of the
incoming hadrons which makes the primary gluon production finite and
preserves unitarity. At sufficiently small momentum fractions $x$ and
small scales $Q^2$, the evolution dynamics of gluon distributions
(BFKL \cite{BFKL} or DGLAP \cite{DGLAP}) is expected to become
dominated by gluon fusions rather than splittings, inhibiting the growth of
the gluon distributions towards smaller $x$ or higher $Q^2$ and
leading to saturation. The idea has been successfully tested in the
case of the free proton: geometric scaling of the structure function
$F_2^p$ at small $x$ and small $Q^2$ predicted by the dipole picture
\cite{GEOMSCALING} has been observed. Also the leading-order (LO) DGLAP fits
to the HERA data in the region $x\sim 10^{-5}$ and $Q^2\sim
1.5$~GeV$^2$ improve after including nonlinearities \cite{GLR,MQ}
induced by the gluon fusions \cite{EHKQS}. Extrapolation to nuclei,
however, is more model-dependent, see e.g. \cite{EHKQS_HP,ARMESTO}.

\subsubsection{Saturation in $A$+$A$ - semianalytical models}

In the context of ultrarelativistic heavy ion collisions, the idea of
gluon saturation in the wave functions of the colliding nuclei
governing and regulating the final state particle production was
discussed first in \cite{BM87}. It was suggested that particle
production in high energy $A$+$A$ collisions could be computed in
perturbative QCD (pQCD) as the saturation scale $Q_{\rm sat}\propto
A^{1/6}$ becomes large, $Q_{\rm sat}\gg\Lambda_{\rm QCD}$. More
recently, initial gluon production in $A$+$A$ collisions was computed
semi-analytically in Refs. \cite{KL,KLN} by applying $k_T$
factorization to $2\rightarrow 1$ processes. In this approach, the
shape of final state rapidity distributions is taken to be directly
that of initially produced gluons, and the overall normalization is
fixed on the basis of RHIC data at one energy. Good  fits to the
multiplicities and pseudorapidity distributions at RHIC have  been
obtained \cite{KL} and predictions for the LHC have been made
\cite{KLN}. It is, however, useful to recall the theoretical
uncertainties of this approach. Widening of the rapidity distributions
due to secondary collisions, as demonstrated by longitudinally
boost-noninvariant hydrodynamics
\cite{EKR97,HIRANO_rap,Sollfrank_prc}, is neglected. This phenomenon
affects off-central rapidities more than the central ones. There are
also uncertainties in the determination of the saturation region for
nuclei (see also \cite{EHKQS_HP}) as well as in the specific form of
the unintegrated nuclear gluon distributions in the saturation region
and also at large $x$ (see the discussion in \cite{EKRT2}).

\subsubsection{Colour glass condensate model for $A$+$A$ - a lattice approach}

The idea of saturation in ultrarelativistic heavy ion collisions was
exploited also in the  colour-glass condensate (CGC) model
\cite{McLV}, where the small-$x$ gluons are described by classical
gauge fields. Over the recent years this model has experienced a very
active stage of development, for recent reviews see
e.g. \cite{CGCreviews}. With boost-invariance the CGC model becomes an
effective 3-dimensional theory. Primary gluon production can then be
computed nonperturbatively by assuming a Gaussian distribution of
random colour charges and solving the classical Yang-Mills equations
on a lattice. For the SU(3) case, this was first done in
\cite{NKV}. The too large transverse momentum relative to the
multiplicity obtained from the calculation was, however, a puzzle (see
e.g. \cite{KJE}) which was solved only recently in \cite{LAPPI}. The
lattice approach, however, does not provide a value for the saturation
scale, and thus no absolute {\em prediction} for the initial state,
either. As discussed in \cite{NKV,LAPPI}, the predicted initial state
of matter can range between  {\em (i)} a dilute system of gluons,
which only fragment into more gluons but do not interact with each
other (as in the scenario of Ref. \cite{KL} discussed above), and,
{\em (ii)} a very dense system where gluons interact vividly and where
the initially produced multiplicity is essentially the final
multiplicity but transverse energy is clearly larger than the final
$E_T$. The latter case is equivalent to what is  predicted in the
pQCD+saturation+hydrodynamics model \cite{EKRT,ERRT} discussed below.
The nonperturbative CGC model \cite{LAPPI} or the semi-analytical
$k_T$ factorization model \cite{KL,KLN} can be used to predict the
initial state at the LHC if the saturation scales and unintegrated
gluon densities can be extracted and the form of colour density
fluctuations constrained on the basis of deep inelastic $F_2^A$ data
for nuclei \cite{RUMMUKAINEN,ASW04}.

\subsubsection{pQCD+final state saturation model in $A$+$A$}

In the pQCD+saturation model \cite{EKRT}, a dense initial
state with large transverse energy is predicted by requiring
saturation among the {\em produced} gluons \cite{EKRT,EKT} in the
same spirit as was originally suggested \cite{GLR,BM87} for the initial
state.  Parton production in a midrapidity unit $\Delta y$ in a
central $A$+$A$ collision is computed by applying collinearly factorized
pQCD cross sections with nuclear parton distributions obtained from a
global DGLAP fit \cite{EKS98} to nuclear deep inelastic scattering
and Drell-Yan data. The parton production is allowed to continue from
large $p_T$ partonic jets, where pQCD is known to work, down to a
semi-hard transverse momentum cut-off $p_0\gg\Lambda_{\rm QCD}$.
If the production is abundant enough the cut-off $p_0$ can be defined
from saturation as follows: let the number of produced partons
(gluons clearly dominate) at $p_T\ge p_0$ in a central ({\bf b} $=$
{\bf 0}) collision be $N_{AA}({\bf 0}, p_0, \Delta y,\sqrt s)$ and the
effective transverse area occupied by each parton to be proportional
to $\pi/p_0^2$. At saturation the partons fill the whole
transverse overlap area $\pi R_A^2$, and the effective saturation
scale $p_0 = p_{\rm sat}$ is a solution of a saturation criterion
\begin{equation}\label{satcrit}
N_{AA}({\bf 0}, p_0, \Delta y=1,\sqrt s)\cdot \pi/p_0^2
  = \pi R_A^2 .
\end{equation}
We have taken the possible constant of proportionality in front of
$\pi/p_0^2$, which might contain e.g. $\alpha_s$ or group
theoretical colour factors, equal to one.

Gluons with $p_T\le p_{\rm sat}$ form at later times,
$\tau>1/p_{\rm sat}$, and they fuse together with gluons
formed earlier. In this sense, the number of produced partons
saturates. The number of partons at saturation, $N_{AA}(\psat)$, gives a
fair estimate of the  total multiplicity of produced partons, provided
that $p_{\rm sat}\gg\Lambda_{\rm QCD}$. The formation time of the
dominant part of the system is then $\tau_0\sim 1/p_{\rm sat}$. With
the saturation condition of Eq.~(\ref{satcrit}), the computation of
the initial state is closed and can be done at any energy. Total
transverse energy deposit and initial energy density at $\tau_0$ can
be computed including the next-to-leading order contributions in
minijet production \cite{ET01}. Benefits over the other approaches
discussed above, are that chemical decomposition and especially the
net-baryon number content of the initial state are also obtained
\cite{EK97}.

The pQCD + (final state) saturation approach obviously does not
suggest a microscopic dynamical mechanism for saturation, although
the general idea of produced gluons to act as a medium for the
to-be-produced gluons, is similar to the self-screening suggested in
\cite{Eskola:1995bp}. The pQCD+saturation model, rather, serves as a useful
effective tool in estimating the magnitude of initial parton
production in simplest situations where only one scale dominates the
production, as in central and nearly central collisions at
midrapidities. In spite of the possibly large theoretical
uncertainties, the pQCD+saturation model, combined with hydrodynamics,
i.e. isentropic spacetime evolution, correctly {\em predicted} the
multiplicities measured in nearly central Au+Au collisions at RHIC at
$\sqrt s=56$, 130 and 200 GeV \cite{EKRT,ERRT,KJE_PVR_B}.

Also the low-$p_T$ spectra of pions, kaons and protons measured in 5\%
central Au+Au collisions at $\sqrt s=130$~GeV at RHIC are
simultaneously reproduced surprisingly  well \cite{ENRR,ENRR_QM02} by
pQCD+saturation+hydrodynamics, provided that the hydrodynamic evolution
decouples at rather high temperature, $T_{\rm dec}\approx 150$~MeV.
Such a high decoupling temperature has also been suggested in a
fireball parametrization \cite{FLORKOWSKI}. Motivated by the success
of the  pQCD+saturation+hydrodynamics model, we here extend the study
of Ref.~\cite{ENRR} to $\sqrt s=200$~GeV at RHIC and $\sqrt
s=5500$~GeV at the LHC as well.

\subsection{Initial densities from pQCD + saturation}\label{sec:inicond}

According to collinear factorization and LO pQCD, the
inclusive cross section of two-parton production in $A$+$B$ collisions is
given by
\begin{equation}
\frac{d\sigma}{dp_T^2dy_1dy_2}^{\hspace{-0.6cm}AB\rightarrow kl+X} =
\sum_{ij}x_1 f_{i/A}(x_1,Q^2) x_2f_{j/B}(x_2,Q^2) \frac{d\hat\sigma}{d\hat
t}^{ij\rightarrow kl}\hspace{-0.6cm}(\hat s,\hat t,\hat u),
\label{2parton}
\end{equation}
where $f_{i/A}$ and $f_{j/B}$ are the number densities of partons $i$
and $j$ in the beams $A$ and $B$, and $x_{1,2}$ are the fractional momenta
of $i$ and $j$.  The hatted symbols are the Mandelstam variables of
the partonic $2\rightarrow2$ subprocesses, whose cross section
$\hat\sigma\propto\alpha_s^2$. The partons $k$ and $l$ are produced
back-to-back in transverse plane, with transverse momentum $p_T$ for
each, and rapidities $y_1$ and $y_2$. We choose the factorization and
renormalization scales to be identical, and set $Q=p_T$. In certain
cases, such as jet production \cite{Ellis:1990ek} and minijet
transverse energy production \cite{ET01}, the computation can be
rigorously extended to next-to-leading order (NLO) pQCD. In these
cases, a $K$ factor which simulates the NLO contributions, can be
defined and determined.

Consider now only partons above a minimum scale
$p_0\gg\Lambda_{\rm QCD}$ where the pQCD computation is still
trustworthy. Using Eq.~(\ref{2parton}), the number $(N^f)$ and
transverse energy $(E_T^f)$ distributions for partons of flavour $f$
falling into a rapidity interval $\Delta y$ can be
defined as \cite{EKL89,EK97}
\begin{eqnarray}
\frac{d\sigma}{dX^f}\bigg|_{y_f\in \Delta y\atop p_T \ge p_0}
&=&\int dp_T^2 dy_1 dy_2 \sum_{\langle kl \rangle}
\frac{d\sigma}{dp_T^2dy_1dy_2}^{\hspace{-0.6cm}AB\rightarrow kl+X}
\frac{1}{1+\delta_{kl}} \\
&&\delta\left(X^f-p_T^{n(X^f)}\left[\delta_{fk}\theta(y_1\in\Delta y)
+\delta_{fl}\theta(y_2\in\Delta y)\right]\right)\theta(p_T\ge p_0),
\nonumber
\end{eqnarray}
where the power $n(X^f)=0$ when $X^f = N^f$, and $n(X^f)=1$ when $X^f = E_T^f$.
The first momenta of these distributions become
\begin{eqnarray}
\sigma\langle X^f \rangle_{\Delta y, p_0}
&=&\int dp_T^2 dy_1 dy_2 \sum_{ij \atop \langle kl \rangle}
x_1 f_{i/A}(x_1,Q^2) x_2f_{j/B}(x_2,Q^2)
\frac{1}{1+\delta_{kl}} \label{sigmaX} \\
&&\left[\delta_{fk} \frac{d\hat\sigma}{d\hat t}^{ij\rightarrow kl}
\hspace{-0.6cm}(\hat s,\hat t,\hat u) +
\delta_{fl} \frac{d\hat\sigma}{d\hat t}^{ij\rightarrow kl}
\hspace{-0.6cm}(\hat s,\hat u,\hat t)
\right]p_T^{n(X^f)} \theta(y_1\in\Delta y)\,\theta(p_T\ge p_0).
\nonumber
\end{eqnarray}

For an $A$+$B$ collision at an impact parameter {\bf b}, the average
number of partons produced with $y\in \Delta y$ and $p_T\ge p_0$ can
then be obtained as
\begin{equation}
N^f_{AB}({\bf b}, p_0, \Delta y, \sqrt s)
= T_{AB}({\bf b}) \sigma\langle N^f \rangle_{\Delta y, p_0}
\label{N_AB^f}
\end{equation}
and their transverse energy as
\begin{equation}
E_{T,AB}^f({\bf b}, p_0, \Delta y, \sqrt s)
= T_{AB}({\bf b}) \sigma\langle E_T^f \rangle_{\Delta y, p_0}.
\label{ET_AB^f}
\end{equation}
The nuclear overlap function,
\begin{equation}\label{TAA}
T_{AB}({\bf b})=\int d^2{\bf r} \ T_A({\bf b}-{\bf r})T_B({\bf r}),
\label{TAB}
\end{equation}
where ${\bf r}=(x,y)$, is defined as usual, through the thickness function
$
T_A({\bf r})=\int dz\rho_A(z,{\bf r})
$,
where $\rho_A(z,{\bf r})$ is the nuclear density. We shall use the
spherically symmetric Woods-Saxon nuclear density \cite{BohrMottelson}.
Below, the flavour-summed quantities ($N$ and $E_T$) appear without the
index $f$. The net-baryon number carried into the rapidity acceptance
$\Delta y$ by the $p_T\ge p_0$ minijets is then one third of the
produced net quark number \cite{EK97},
\begin{eqnarray}\label{eq:netbaryons}
N_{AB}^{B-\overline B}({\bf b}, p_0, \Delta y, \sqrt s)
& = & T_{AB}({\bf b}) \, \frac{1}{3} \sum _q
\left[
\sigma\langle N^q \rangle_{\Delta y, p_0} -
\sigma\langle N^{\bar q} \rangle_{\Delta y, p_0}
\right] \\ \nonumber
& \equiv & T_{AB}({\bf b})
\sigma\langle N^{B-\overline B} \rangle_{\Delta y, p_0}.
\end{eqnarray}

As discussed in the previous section, the dominant transverse scale in
primary production, $p_0=p_{\rm sat}$, in central rapidities of
ultrarelativistic  $A$+$A$ collisions is obtained by solving
Eq.~(\ref{satcrit}). Since the predictions of multiplicities at RHIC
agreed with the measurements, we do not further tune the initial
state computation from that in Ref. \cite{ERRT}, i.e. we use the
GRV94 parton densities \cite{GRV94} together with nuclear effects from
the EKS98 parametrization \cite{EKS98}, and $K$ factors 1.6 (2.3) for
RHIC (LHC) based on the NLO analysis of minijet $E_T$ production
\cite{ET01}. Note that the $K$-factors have been shown to vary with the
PDFs in a manner that the changes in $\sigma\langle E_T\rangle$ remain small even if
newer PDFs were applied \cite{ET01}.

Table~\ref{tab:pQCD} shows the saturation scales, formation times and
the pQCD quantities computed at $p_0=p_{\rm sat}$ for RHIC and LHC
in central and nearly central $A$+$A$ collisions. The centrality
selection is simulated as in Ref. \cite{ERRT}, by considering a
central collision of an effective nucleus $A_{\rm eff}<A$: first,
e.g. for 5~\% most central collisions, we find a maximum impact
parameter by taking 5~\% of the $A$+$A$ total cross section in a
(optical) Glauber formulation.  Then, we determine $A_{\rm eff}$ by
requiring that the number of participants in a central
$A_{\rm eff}$+$A_{\rm eff}$ collision equals the average number of
participants computed for the 5~\% central $A$+$A$ collisions.

%%%%%%%%%%%%%%%%%%%%% BEGIN TABLE %%%%%%%%%%%%%%%%%%%%%%%%%%%%%%%%	
\begin{table}[h]
\begin{center}
\begin{tabular}{|c||c|c||c|c|c|c||c|c|}
\hline
$\sqrt s/A$ [GeV] & \multicolumn{2}{c||}{130} & \multicolumn{4}{c||}{200}
              & \multicolumn{2}{c|}{5500} \\
\hline
Centrality 			& 0 \% 	& 5 \% 	& 0 \% 	& 5 \% 	& 10 \%	& 15 \%	& 0 \% 	& 5 \% \\
\hline
\hline
$A,A_{\rm eff}$ 	& 197 	& 181 	& 197 	& 181 	& 166 	& 153	& 208 	& 193 \\
$T_{A_{\rm eff}A_{\rm eff}}(0)$ [1/mb]
				 	& 29.4	& 26.1	& 29.4	& 26.1	& 23.1	& 20.6	& 31.7 	& 28.5\\
$N_{\rm part}$ 		& 377 	& 346 	& 379 	& 347 	& 318 	& 293	& 407	& 376 \\
$p_{\rm sat}$ [GeV]	& 1.08 	& 1.06 	& 1.16 	& 1.15 	& 1.14 	& 1.12	& 2.03 	& 2.01 \\
$\tau_0$ [fm/c] 	& 0.18 	& 0.19 	& 0.17 	& 0.17 	& 0.17 	& 0.18	& 0.10 	& 0.10 \\
$\sigma\langle E_T\rangle$ [mbGeV] 
					& 65.1 	& 67.0 	& 83.6 	& 85.9 	& 88.3 	& 90.7	& 468 	& 479 \\
$\sigma\langle N^g\rangle$ [mb]    
					& 35.9 	& 37.3 	& 42.9 	& 44.6 	& 46.4 	& 48.2	& 135 	& 140 \\
$\sum_q\sigma\langle N^q\rangle$ [mb] 
					& 4.50 	& 4.66 	& 4.43 	& 4.58 	& 4.75 	& 4.92	& 6.46 	& 6.66 \\
$\sum_q\sigma\langle N^{\bar q}\rangle$ [mb] 
					& 2.65 	& 2.76 	& 2.87 	& 2.98 	& 3.10 	& 3.21	& 6.14 	& 6.34 \\
$\sigma\langle N\rangle$ [mb]    
					& 43.1 	& 44.7 	& 50.2 	& 52.2 	& 54.3 	& 56.3	& 148 	& 153 \\
$\sigma\langle N_{B-\bar B}\rangle$ [mb]    
					& 0.616	& 0.635	& 0.520	& 0.536	& 0.553	& 0.570	& 0.106	& 0.109 \\
\hline
\end{tabular}
\end{center}
\caption{\small The pQCD quantities computed from Eq.~(\ref{sigmaX})
with $p_0=p_{\rm sat}$ in central $A$+$A$ and in central
$A_{\rm eff}$+$A_{\rm eff}$ collisions at RHIC and LHC corresponding to
various centrality cuts and $\Delta y=1$. The saturation scales from
Eq.~(\ref{satcrit}), the corresponding formation times, and also the
number of participants and the overlap function at ${\bf b}={\bf 0}$ are shown.
}
\label{tab:pQCD}
\end{table}
%%%%%%%%%%%%%%%%%%%%% END TABLE %%%%%%%%%%%%%%%%%%%%%%%%%%%%%%%%

The pQCD calculation of minijet production is a momentum space
calculation. In order to define the initial densities, a connection
between the momentum of the minijet and its space-time formation point
is needed. At collider energies the nuclei are strongly contracted,
$2R_A/\gamma_{\rm cm}\ll 1$~fm, and the typical partons at saturation
have a longitudinal spread of $2/(x\sqrt s)\sim 1/p_{\rm sat}\ll
1$~fm. We therefore consider the collision region as a point in the
longitudinal direction and assume that the rapidity of the minijet
coincides with the space-time rapidity of the formation point,
$y=\eta=(1/2)\ln[(t+z)/(t-z)]$. The formation (proper) time we take
to be the inverse of the saturation scale, $\tau_0=1/p_{\rm sat}$.
Thus the minijet matter forms along the hyperbola
$t=\sqrt{z^2+\tau_0^2}$ with initial longitudinal flow velocity
$v_z(\tau_0)=z/t$. The partons in the rapidity interval $\Delta y$
thus occupy a volume $\Delta V(\tau_0)=A_T\Delta z=\pi R_A^2\tau_0\Delta y$
at proper time $\tau_0$.

The average initial energy density, number density and, as a new element
\cite{EK97,ENRR}, the net-baryon number density of the produced parton
matter in central $A$+$A$ collisions are then obtained as
\begin{eqnarray}
\langle \epsilon \rangle
&=& \frac{1}{\Delta V(\tau_0)} E_{T,AA}({\bf 0}, p_{\rm sat}, \Delta
y, \sqrt s)\\
\langle n \rangle
&=& \frac{1}{\Delta V(\tau_0)} N_{AA}({\bf 0}, p_{\rm sat}, \Delta y,
\sqrt s) \\
\langle n_B \rangle
&=& \frac{1}{\Delta V(\tau_0)}N_{AA}^{B-\overline B}({\bf 0},
p_{\rm sat}, \Delta y, \sqrt s).
\end{eqnarray}
For nuclei with realistic transverse profiles, however, the saturation
scale and thus also the formation time may vary with the transverse
coordinate ${\bf r}=(x,y)$ \cite{EKT,KHHET}. We avoid such further
modeling here and, following Ref. \cite{ERRT}, we consider a constant
formation time $\tau_0 = 1/p_{\rm sat}$ and extract the transverse
profile for the initial densities by differentiating the overlap function
$T_{AB}$ with respect to $d^2r$.  The nucleon--nucleon
luminosity for a transverse area element
$d^2r$ is $T_A({\bf b}-{\bf r})T_{B}({\bf r})$ and the
volume element $dV=dz\,d^2r=\tau \Delta y\,d^2r$. These lead to the
${\bf r}$ dependent energy density and net-baryon density at
$(\tau_0=1/p_{\rm sat},\eta = 0)$ in central (${\bf b}={\bf 0}$) $A$+$A$
collisions,
\begin{eqnarray}
\epsilon(\tau_0,{\bf r})
&=& \frac{dE_T}{\tau_0d\eta d^2r}
= T_A({\bf r})T_A({\bf r})
\frac{\sigma\langle E_T \rangle_{\Delta y, p_{\rm sat}}}{\tau_0 \Delta y}
\label{epsilon}
\\
n_B(\tau_0,{\bf r}) &=& \frac{dN_{AA}^{B-\overline B}}{\tau_0d\eta d^2r}
= T_A({\bf r})T_A({\bf r})
\frac{\sigma\langle N^{B-\overline B} \rangle_{\Delta y, p_{\rm sat}}}
{\tau_0 \Delta y} \ .
\label{nB}
\end{eqnarray}

\begin{figure}[tbh]
\begin{minipage}[t]{70mm}
\vspace{-8mm}
    \epsfysize 11.0cm \epsfbox{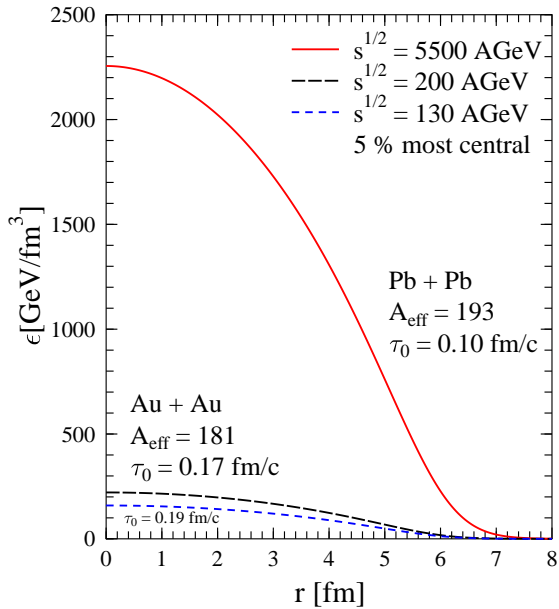}
\end{minipage}
 \hspace{14mm}
\begin{minipage}[t]{70mm}
 \vspace{20mm}
 \begin{center}
  \begin{minipage}{65mm}
\caption{\protect\small
Transverse profile of the initial energy density in an average
$A$+$A$ collision in 5 \% most central collisions at the LHC (solid line)
and RHIC (dashed lines). Formation times and effective nuclei
are indicated, see Table~\ref{tab:pQCD}.
}
  \label{fig:inicond}
  \end{minipage}
 \end{center}
\end{minipage}
\hspace{\fill}
\vspace{-2.2cm}
\end{figure}

In Fig. \ref{fig:inicond} we show the initial energy density in
5 \% most central Au+Au collisions at $\sqrt s = 200$~$A$GeV at RHIC and
Pb+Pb collisions at $\sqrt s = 5500$~$A$GeV at the LHC.
Notice the very large densities in the very center of the system.
It should, however, be emphasized that since the {\em volume} near $r\sim0$
is small, the average energy densities are clearly smaller than
those in the center, see Table 1 of Ref. \cite{EKRT}.
Another interesting point is the large difference between RHIC and LHC.
This is on one hand due to the enhanced gluon production at higher energies
(see Table~\ref{tab:pQCD}) and on the other due to the shorter formation
time (smaller initial QGP volumes) in the LHC case (see Eq.~(\ref{epsilon})).
Since the initial transverse size is nearly the same in both cases, the
radial pressure gradients are larger at the LHC leading to stronger
flow, as discussed in \cite{PH_QM02}.

%%%%%%%%%%%%%%%%%%%%% FIGURE %%%%%%%%%%%%%%%%%%%%%%%%%%%%%%%%
%%%%%%%%%%%%%%%%%%%%% FIGURE %%%%%%%%%%%%%%%%%%%%%%%%%%%%%%%%

The net-baryon number in central rapidity unit,  Eq.~(\ref{eq:netbaryons}), 
can now be obtained from Table~\ref{tab:pQCD}.  For RHIC, we get 
$N_{B-\bar B}=16.6$ and 14.0 in 5 \% most central Au+Au collisions at 
$\ssNN=130$ and 200 GeV, respectively. For the LHC, we predict $N_{B-\bar B}=3.11$ 
in 5 \% most central Pb+Pb collisions at $\ssNN=5500$~GeV.

\section{From dense initial state to decoupled hadrons}\label{sec:hydro}

\subsection{The hydrodynamical framework}

For the initial state of matter, the pQCD+saturation model provides
the initial energy density $\epsilon$ and net-baryon number density
$n_B$ over a transverse profile at the formation time $1/p_{\rm sat}$
of the system. We make a bold assumption of fast
thermalization by taking the thermalization time scale to be the same
as the formation time, $\tau_0=1/p_{\rm sat}$. Some arguments can be
given in favour of this assumption. First, as discussed in
\cite{EKRT,EKT,EKT2}, from the point of view of $\langle \epsilon
\rangle$ and $\langle n \rangle$, the saturated (fully gluonic) system
indeed looks thermal. This suggests that only elastic collisions causing
energy and momentum transfer are needed for thermalization, number-changing 
reactions are not essential. Second, the formation time of
more energetic (mini)jets is shorter than the time scale $\tau_0$
associated with the saturation scale, so some secondary interactions
have taken place already before the overall formation time. Third,
recent studies of plasma instabilities lend support to fast
thermalization \cite{MOORE,Rebhan:2004ur} instead of the slower bottom-up
thermalization \cite{Baier:2000sb}. We also note that we assume full
chemical equilibrium and follow the evolution of net-baryon number,
which as a conserved quantity should not be sensitive on the details
of chemical processes.  We do not discuss the dynamics of chemical
equilibration of the QGP here.  Recent studies on this can be found
in Ref. \cite{NRR}.

\subsubsection{Simplifications: longitudinal scaling flow and
  azimuthal symmetry}

As discussed in Sec. \ref{sec:inicond}, in our minijet description the
parton system forms with an initial longitudinal flow velocity
$\tanh\theta=v_z=z/t$. Assuming that the expansion maintains this
form, the original (3+1)-dimensional problem reduces to
(2+1)-dimensional evolution equations for transverse expansion. For
central collisions the final state is azimuthally symmetric leaving
only the (1+1)-dimensional equations for radial flow to be solved. In
this study we do not consider non-central collisions which exhibit the
interesting phenomena of azimuthally asymmetric flow.

Obviously boost invariance will be broken in real collisions with
finite collision energy.  However, the assumption of boost invariant
flow is still very good in the central rapidity region, as will be
argued next. The effects of longitudinal pressure gradients on
longitudinal velocity were studied in \cite{EKR97}, where the initial
densities were computed from the pQCD minijet rapidity distributions
and the initial longitudinal flow velocity was assumed to satisfy the
scaling form. Since the radial expansion was neglected, the study \cite{EKR97}
overestimates the change of the flow because cooling is slower and
system lives longer than with radial flow. The obtained flow deviated
only marginally (by a few per cent) from the scaling flow at the end of
the evolution in the central rapidity region.  Similar observation
was made in a (2+1)-dimensional analysis at RHIC energy
\cite{HIRANO_rap}.

As the scaling flow does not change much, the transfer of entropy,
deposited initially to the central rapidity unit, to larger rapidities
due to the development of the longitudinal flow is negligible. Also,
the dominant part of the final (thermal) spectra of
hadrons comes from the part of the decoupling surface where
$|y-\theta|\lsim1$, because only a small fraction of particles in
thermal bath have large enough energy to end up in central rapidity
over a rapidity gap larger than one. All in all this means that the
fragmentation regions, where the assumption of boost invariance breaks
down, do not affect significantly the results obtained using scaling
flow at the central rapidity.

The question of a rapidity plateau in the final rapidity density
$dN/dy$ of hadrons is discussed in more details by Hirano et
al.~\cite{HIRANO_rap}. They show that the plateau would appear in
the measured spectra only if the initial densities are flat over a
considerable range in space-time rapidity. At RHIC the measured
pion spectra are Gaussian \cite{Bearden:2001qq} and it remains to be
seen if a plateau develops at LHC energy. In any case, as argued
above, a measured Gaussian rapidity density does not indicate that
the description of transverse flow in the central rapidity region 
based on the assumption of scaling longitudinal velocity is an 
invalid approximation.

We note that the approximations mentioned above can be relaxed in
different ways. There has been a lot of discussion on elliptic
flow, see e.g. a review \cite{Kolb:2003dz}, in which case the
approximation of boost invariant flow is retained but the azimuthal
symmetry is relaxed. Also, as mentioned above, a genuinely 3+1
dimensional hydrodynamical code for collider energies has been
successfully developed \cite{Hirano:2001eu,HIRANO_TSUDA,HIRANO_NARA}.

\subsubsection{Initial transverse flow?}

To complete the initial conditions for the hydrodynamic calculation
the initial collective transverse motion, $v_r(\tau_0,r)$, must be
specified. It can be argued that the change in the number of primary
collisions, when going from the center to the edges in the transverse
plane, could lead to non-zero initial transverse flow. However,
hydrodynamical studies at the SPS energy show that only a very weak
initial flow can be tolerated without getting into disagreement with
the slopes of transverse hadron spectra \cite{Huovinen:2001wx}. 
Since there is no obvious reason for a strong transverse initial flow,
we take for simplicity $v_r(\tau_0,r)=0$.

In Ref.~\cite{Kolb:2002ve} it was found that especially the spectra
of protons, at RHIC, are better reproduced if some initial velocity is
introduced. However, they use $\tau_0=0.6$ fm/c as the thermalization
time and argue that the initial velocity arises from pre-thermal
interactions. The velocity they assume is similar in magnitude to that
created in hydrodynamical evolution up to $\tau_0=0.6$ fm/c if
$\tau_0=1/\psat\sim0.2$ fm/c is assumed, as we do.
Whether or not the system is fully thermalized at $\tau_0=1/\psat$,
this early starting of hydrodynamical evolution can effectively
describe the initial buildup of transverse flow as a result of the
collisions which lead to thermalization.

\subsubsection{Equation of State}

Another key input in the hydrodynamic studies is the Equation of
State, $P=P(\epsilon,n_B,n_S)$, specifying pressure as a
function of energy density, net-baryon number density and net 
strangeness density.  Since
strong interactions conserve strangeness, the total strangeness in
the expanding matter of a heavy ion collision is zero.  In principle
non-zero strangeness spots could be generated in the matter through
fluctuations but we assume that the net strangeness vanishes
locally. This condition is imposed on the EoS and we
write simply $P(\epsilon,n_B)$ for $P(\epsilon,n_B,n_S=0)$.
With the initial conditions considered here, the effects of $n_B$ on
pressure are negligible, which we have checked, and for the 
hydrodynamical simulations we use $P(\epsilon,n_B=0)$ for simplicity. 
However, when calculating the final hadron spectra the value of $n_B$ 
and the condition $n_S=0$ are essential in determining the baryon and 
strangeness chemical potentials which enter the thermal distributions 
of hadrons (see Eqs.~(\ref{CooperFrye}) and (\ref{strangeness}) ahead). 
When hydrodynamic simulations are converted to hadron spectra,
the EoS in full $(\epsilon,n_B)$-plane is used.

In principle the EoS can be obtained from lattice QCD.  A full EoS
with finite quark masses and finite baryochemical potential $\mu_B$
is, however, not yet available. It is expected that for
small values of $\mu_B$ relevant at central rapidity region at
collider energies the transition from hadron gas to parton matter is
a rapid crossover \cite{Rajagopal:2000uu}. For the recent progress
on lattice calculations, see e.g. \cite{Karsch:2004wd,Karsch:2003jg}.
From the point of view of radial hydrodynamic flow the difference 
between a weak first order transition and a rapid cross over is 
qualitatively not very significant as long as the EoSs are similar 
away from the transition region. Since we would like to be able to 
include the dependence of spectra on quantum numbers through the 
chemical potentials, we use in calculations an EoS with ideal
quark-gluon plasma (QGP) in the high temperature phase and a hadron
resonance gas (HRG) below the transition. For these phases it is
easy to include the chemical potentials into the EoS.

A first order transition is implemented by introducing a bag
constant $B$ into the QGP phase and connecting the two phases via
the Maxwell construction. The inclusion of resonances in the hadron
phase mimics the effects of both attractive and repulsive
interactions between hadrons reasonably well~\cite{Raju}.
We include all hadron and hadron resonance states up to 
$m=2$~GeV. A detailed account of constructing such an EoS can be found
e.g. in Ref.~\cite{Sollfrank_prc}. With $N_f=3$ we set the
bag constant $B$ to be $B^{1/4}=239$~MeV, giving $T_c=165$ MeV 
for the transition temperature. 

Another way to construct an EoS would be to parametrize the 
$n_B=0$ lattice results above the QCD phase transition \cite{Schneider:2001nf} 
and join it smoothly with HRG EoS at $T_c$. We have checked that
hadron $p_T$ spectra obtained with such an EoS do not differ significantly 
from those obtained with the bag model QGP--HRG 
EoS\footnote{For a recent study on the EoS effects 
on elliptic flow, see \cite{Huovinen:2005gy}.}. 
In this paper, we shall show results with the QGP--HRG EoS only.

\subsubsection{Hydrodynamic equations}

With the longitudinal scaling flow and the assumed azimuthal
symmetry, the transverse velocity and thermodynamic densities become
functions of proper time $\tau$ and transverse coordinate
$r$ only: $v_r=v_r(\tau,r)$, $\epsilon=\epsilon(\tau,r)$,
$n_B=n_B(\tau,r)$, $P=P(\tau,r)$,~\ldots.  The hydrodynamic equations
are
\begin{equation}\label{eq:hydro}
\partial_{\mu}T^{\mu\nu}(x) = 0
\quad\quad {\rm and}\quad\quad \partial_{\mu}j_B^{\mu}(x)=0,
\end{equation}
where
$$
T^{\mu\nu}=(\epsilon + P)u^{\mu}u^{\nu} -Pg^{\mu\nu}
\quad\quad {\rm and}\quad\quad j_B^{\mu}=n_Bu^{\mu}
$$
are the stress-energy tensor and the net-baryon current.
With the above assumptions equations (\ref{eq:hydro}) simplify
to 1+1 dimensional forms \cite{KATAJA,Ruuskanen:1986py}. Given an
EoS in the form $P=P(\epsilon,n_B=0)$ with $n_S=0$ as discussed above,
they can be easily solved numerically.

\subsubsection{Decoupling and decays}

In hydrodynamic description the final state hadron spectra are
obtained by folding the thermal motion with the fluid motion on a
decoupling surface $\sigma(\tau,r)$, which we define from the
condition $T(\tau,r)=\Tdec$. Given the velocity field and the
thermodynamic densities, we calculate the thermal spectra of hadrons
$h$ by using the Cooper and Frye prescription \cite{COOPER_FRYE}
\bea
\pi{dN^h\over d^3{\bf p}\slash E} &=& {dN^h\over dydp_T^2} =
\pi\int_\sigma d\sigma_{\mu}(x)p^{\mu}f_h(x,p;T(x),\mu_h(x)) \\
 &=& {g_h\over 2\pi}
  \sum_{n=1}^\infty (\pm 1)^{n+1} \int_\sigma  r\tau
  e^{n\mu_h(\tau,r)/T(\tau,r)} \label{CooperFrye} \\
  &\big[&\!\!\!\!-p_T I_1(n \gamma_rv_r{p_T\over T})
  K_0(n\gamma_r {m_T^h\over T})\,d\tau
  +m_T^h I_0(n\gamma_rv_r{p_T\over T})K_1(n\gamma_r{m_T^h\over T})\,dr\big]\,,
  \nonumber
\eea
where $\mu_h=B_h\mu_B+S_h\mu_S$ is the chemical potential, $g_h$ the
spin degeneracy factor and $m_T^h=\sqrt{m_h^2+p_T^2}$ the transverse
mass of the hadron of type $h$.  The last form,
Eq.~(\ref{CooperFrye}), holds for the boost invariant cylindrically
symmetric radial flow. The upper (lower) sign is for mesons (baryons)
and $\gamma_r=(1-v_r^2)^{-1/2}$.  Also, the condition
$\mu_h(\tau,r)<m_h$ must always be satisfied for mesons, and,
for the expansion in Eq.~(\ref{CooperFrye}) to hold, also for baryons.\footnote{For %%@
$\mu_h(\tau,r)>m_h$,
regions with $\mu_h(\tau,r)$ bigger or smaller than $m_T^h$ must be
treated separately.} This is the case here.

Chemical potentials for baryon number and strangeness, $\mu_B$ and
$\mu_S$, are obtained on the decoupling surface by expressing the
calculated net-baryon density and the strangeness neutrality in terms
of free-particle densities:
\begin{equation}\label{strangeness}
n_B = \sum_h B_h n_h(T,\mu_h) \quad\quad {\rm and}\quad\quad
  0 = \sum_h S_h n_h(T,\mu_h),
\end{equation}
where $B_h$ is the baryon number and $S_h$ the strangeness of a hadron $h$.

The preferred value $T_{\rm dec}=150$~MeV is extracted on the basis of
RHIC data \cite{ENRR} but we will also show the sensitivity  of our
results to $T_{\rm dec}$. The thermal spectra are calculated for all
hadrons and resonances included in the hadronic EoS.

To obtain final spectra which can be compared with the experimentally
observable ones, we consider all strong and electromagnetic two- and
three body decays of hadron resonances, with branching ratios obtained
from \cite{partdata}. Decay kinematics is discussed in details e.g. in
\cite{kinematics}. Also feed-down effects from weak decays of hyperons
will be discussed.

\subsection{Properties and effects of collective motion}

\subsubsection{Longitudinal flow and transfer $E_T$}

The effects of collective motion arise from the thrust due to
collisions among produced particles, described by pressure in the
hydrodynamic treatment. The most important overall effect of the
rapid longitudinal expansion is to transfer a large fraction of
(transverse) energy from central region to longitudinal motion, see
Ref. \cite{EKR97} for a discussion. The measured final transverse
energy at RHIC is roughly a third of what is predicted by
pQCD+saturation, and the same is expected to happen at the LHC as
well \cite{ERRT}. At the same time, however, the number of
calculated minijets is quite close to the final multiplicity.  In
the hydrodynamic approach the approximate conservation of the number
of particles is a consequence of the (approximate) conservation of
entropy. The initial energy density, or equivalently, since instant
thermalization is assumed, the initial entropy density obtained from
pQCD+saturation thus essentially determines the final multiplicity
\cite{EKRT}.

Longitudinal expansion alone would cool and dilute the matter to free
hadrons. However, with the very large initial densities created at
collider energies, the time needed for this, the longitudinal time
scale, is so long that effects in the transverse direction cannot be
neglected. From the point of view of the transverse energy at central
rapidities we have two opposing effects: In the longitudinal expansion
the large initial thermal energy is transferred to longitudinal motion
and the transverse energy is reduced. On the other hand, the build-up
of transverse flow limits this transfer and part of the initial
thermal energy is converted to the transverse collective motion and
thus to the measured transverse energy. Even though the fraction of
energy transfer into transverse collective motion is much less than
into longitudinal, it produces an effect which is seen as an increase
in the average transverse momentum (see e.g. \cite{measured-pav}) 
or in the effective temperature, measured as the inverse of the slope
of the spectra.

\subsubsection{Properties of transverse flow}

\begin{figure}[!p]
\vspace{-0.5cm}
\hspace{-0.5cm}
   \epsfysize 9.0cm \epsfbox{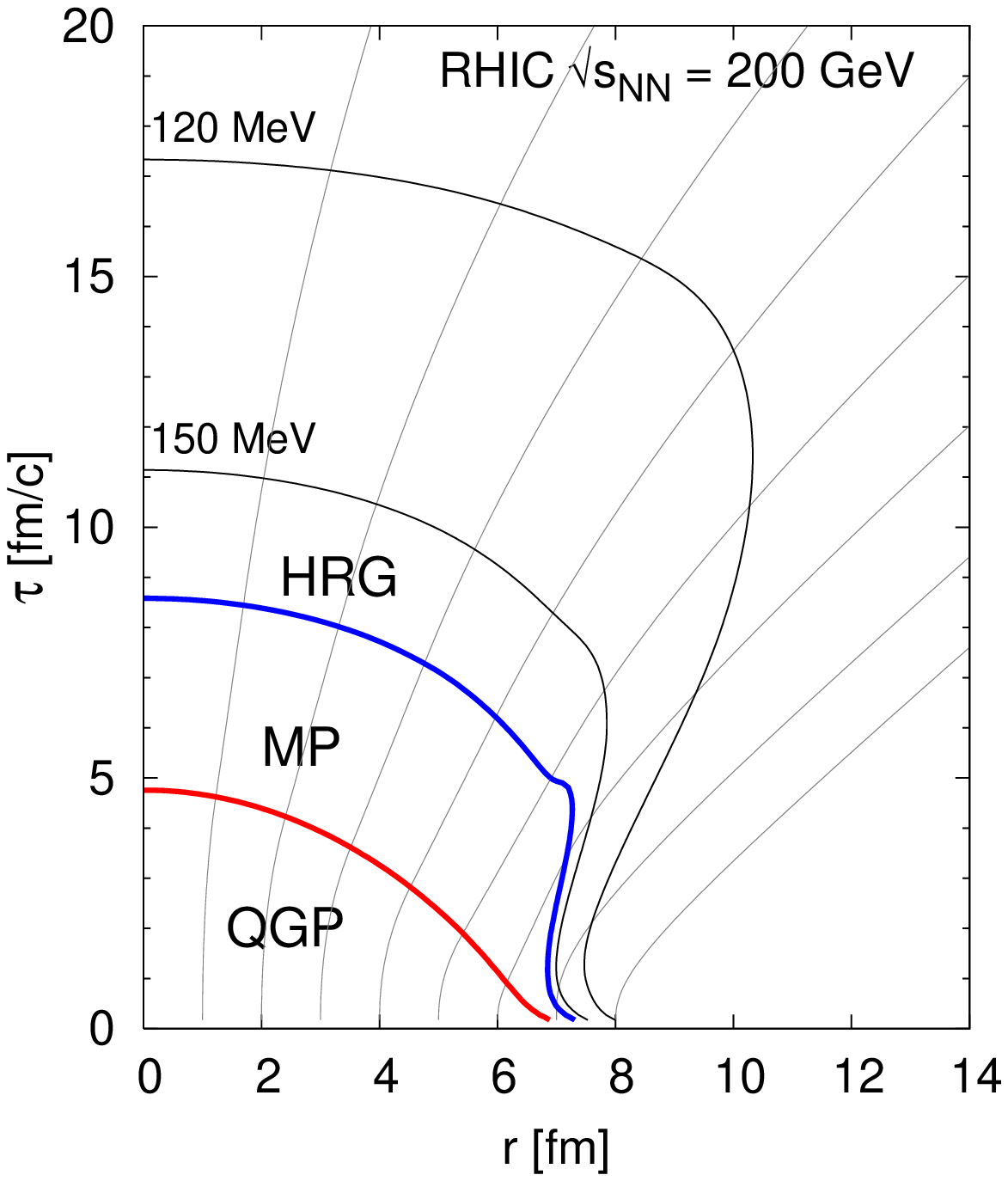} \hspace{0.0cm}
   \epsfysize 9.0cm \epsfbox{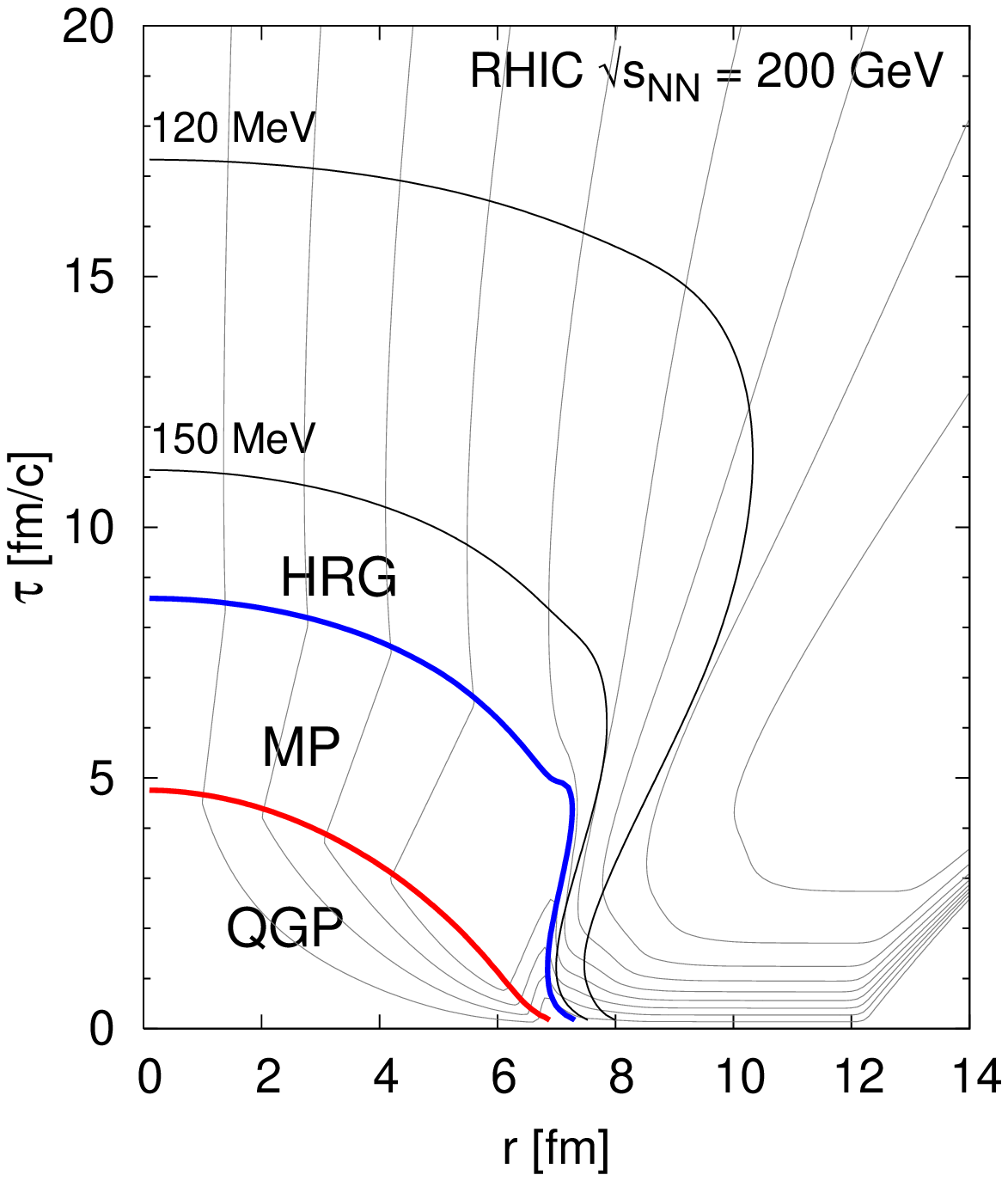}

\vspace{-0.5cm}
\begin{minipage}[t]{75mm}
\caption[a]{\protect\small
The phase boundaries and the two decoupling curves studied, $T_{\rm
dec} = 150$ and $120$ MeV, with selected flowlines in the plane of
proper time $\tau$ and transverse variable $r$ in 5\% most central
Au+Au collisions at $\sqrt{s_{NN}}=200$ GeV.
}
\label{fig:flowRHICa}
\end{minipage} 
\hfill
\begin{minipage}[t]{75mm}
\caption{\protect\small
Contours of transverse velocity $v_r$ in the $(\tau, r)$-plane in 5\%
most central Au+Au collisions at $\sqrt{s_{NN}}=200$ GeV.  Reading
from left to right, $v_r = 0.1,\, 0.2, \dots, 0.9$. 
} 
\label{fig:flowRHICb}
\end{minipage} 
\end{figure}

\begin{figure}[!p]
\vspace{-0.5cm}
\hspace{-0.5cm}
   \epsfysize 9.0cm \epsfbox{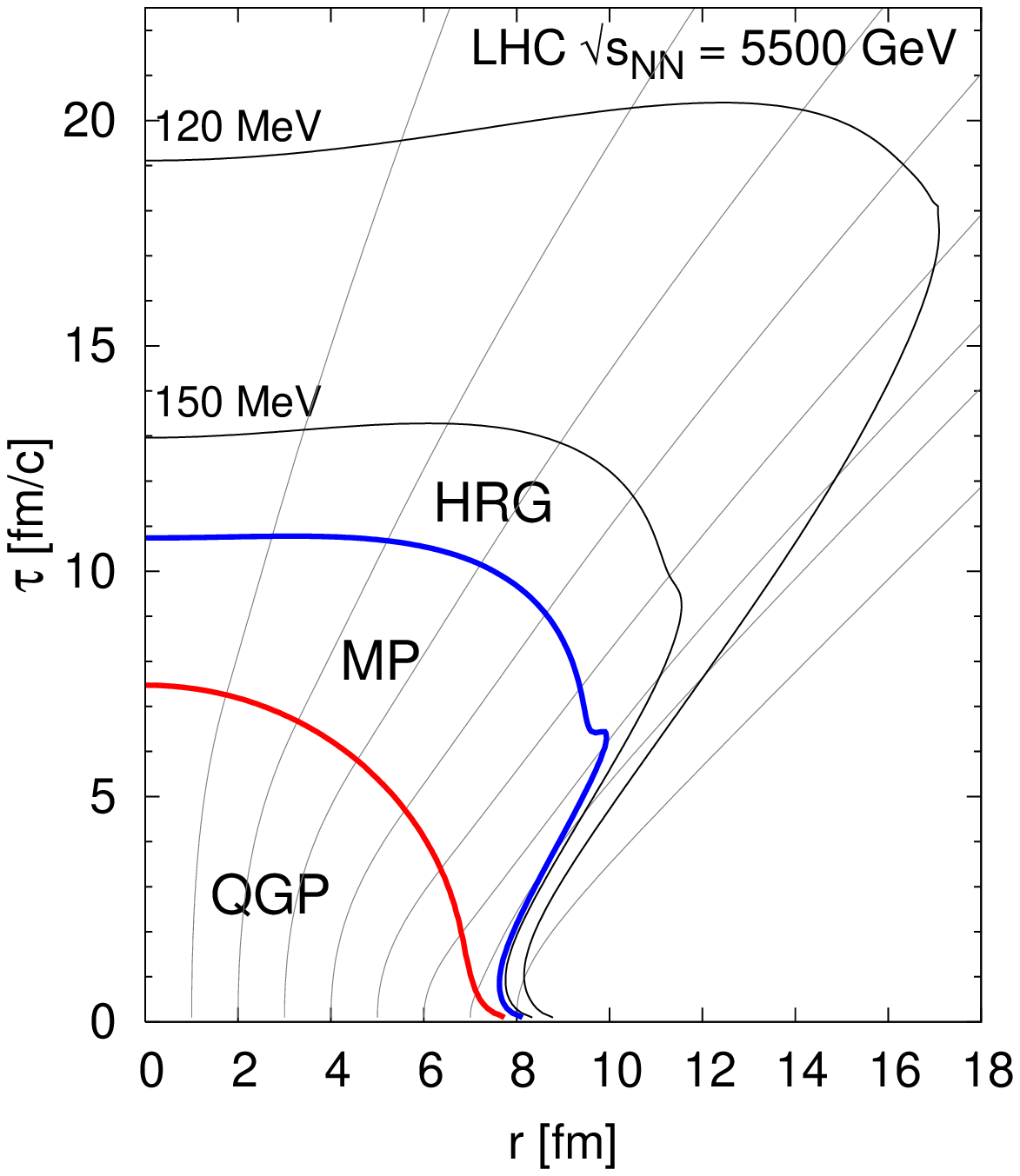} \hspace{0.0cm}
   \epsfysize 9.0cm \epsfbox{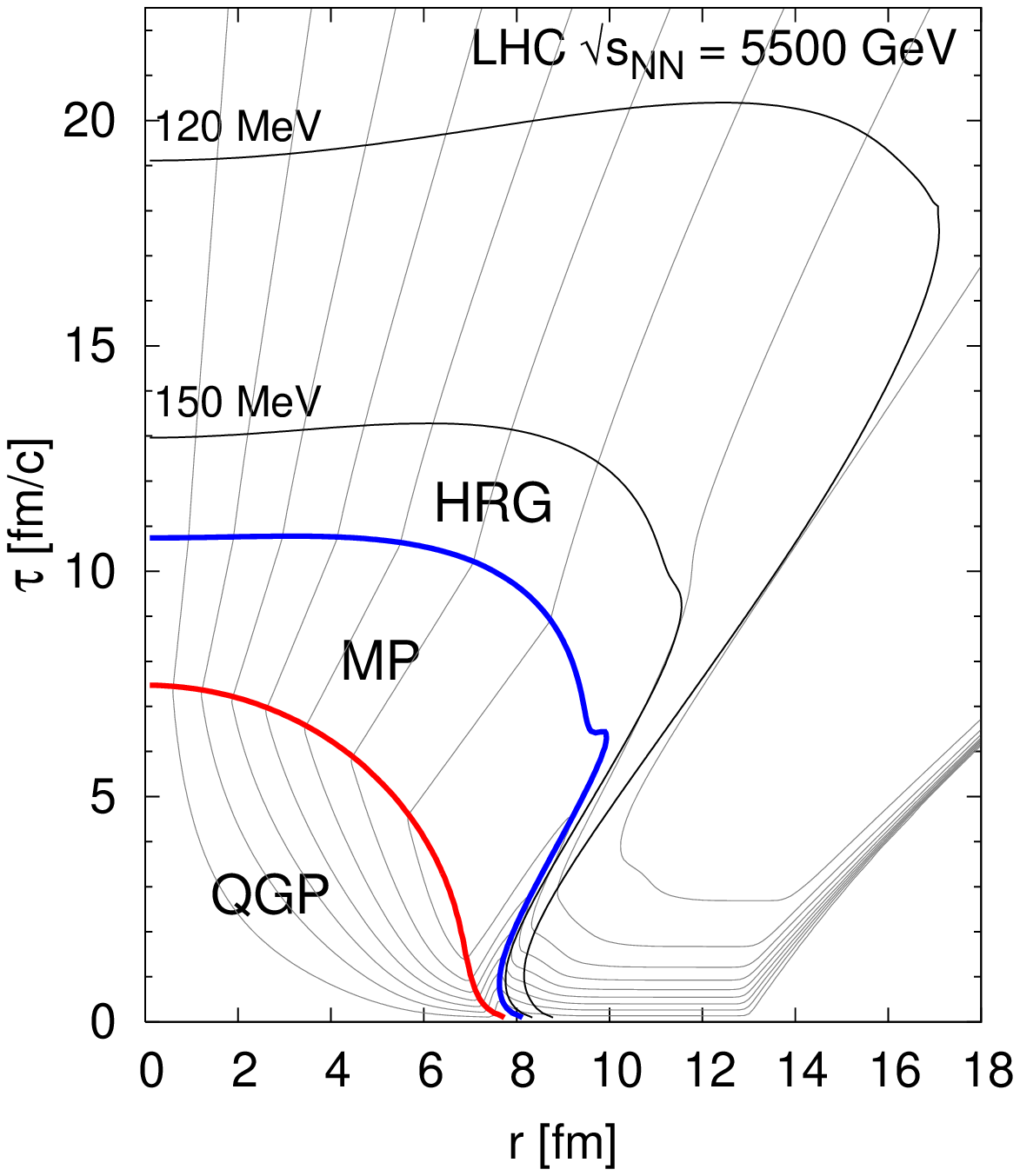}

\vspace{-0.cm}
\begin{minipage}[h]{75mm}
\caption[a]{\protect\small
As Fig.~\ref{fig:flowRHICa} but for 5\% most central Pb+Pb collision
at $\sqrt{s_{NN}}=5500$ GeV.
}
\label{fig:flowLHCa}
\end{minipage} 
\hfill
\begin{minipage}[h]{75mm}
\caption{\protect\small
As Fig.~\ref{fig:flowRHICb} but for 5\% most central Pb+Pb collision
at $\sqrt{s_{NN}}=5500$ GeV.
}
\label{fig:flowLHCb}
\end{minipage} 
\end{figure}

The main overall effect of transverse flow in the hydrodynamic evolution
is a more rapid cooling, leading to clearly shorter lifetime of the
thermal system than with the longitudinal expansion alone. This is
illustrated in Figs.~\ref{fig:flowRHICa}--\ref{fig:flowLHCb}
which show our results for the hydrodynamic spacetime evolution
in the framework discussed above. The figures \ref{fig:flowRHICa} 
and \ref{fig:flowRHICb} are for RHIC and
the next two for the LHC. The figures \ref{fig:flowRHICa} and \ref{fig:flowLHCa}
show the flow lines in the $(\tau,r)$ plane, and the figures \ref{fig:flowRHICb} and %%@
\ref{fig:flowLHCb} the transverse velocity
contours. The phase boundaries QGP--mixed phase (MP) and MP--HRG are
shown by the thick curves. Also drawn are two isotherms corresponding
to $T_{\rm dec}=150$~MeV and $T_{\rm dec}=120$~MeV.

%%%%%%%%%%%%%%%%%%%%% FIGURE %%%%%%%%%%%%%%%%%%%%%%%%%%%%%%%%
%%%%%%%%%%%%%%%%%%%%% FIGURE %%%%%%%%%%%%%%%%%%%%%%%%%%%%%%%%
%
%%%%%%%%%%%%%%%%%%%%% FIGURE %%%%%%%%%%%%%%%%%%%%%%%%%%%%%%%%
%%%%%%%%%%%%%%%%%%%%% FIGURE %%%%%%%%%%%%%%%%%%%%%%%%%%%%%%%%

It is interesting to compare the QGP lifetime at $r=0$ with that in a
system subject to longitudinal scaling flow only. Such a
system would cool down from $\epsilon_0$ to
the critical energy density of QGP, $\epsilon_c=1.93$~GeVfm$^{-3}$ at
$\tau_c=\tau_0(\epsilon_0/\epsilon_c)^{3/4}$. For the maximum values
from the Fig. \ref{fig:inicond}, $\epsilon_0 \approx 220$~GeVfm$^{-3}$ at
RHIC and 2260~GeVfm$^{-3}$ at the LHC, and for the initial times
from Table~\ref{tab:pQCD}, we get $\tau_c=6.0$~fm/c at RHIC and
$\tau_c=20$~fm/c at the LHC. These are to be contrasted with
$\tau_c(r=0)\approx 5$~fm/c at RHIC and 7.5~fm/c at the LHC in Figs.
\ref{fig:flowRHICa}\ldots\ref{fig:flowLHCb}. The effects of transverse
expansion thus become important already at the QGP phase at the LHC.

The evolution of the transverse flow is clearly seen in these
figures. The flow lines $r=r(t)$ (at $z=0$, $\tau=t$) 
in Figs.~\ref{fig:flowRHICa} and \ref{fig:flowLHCa}
depict how a fluid element in a given differential interval in
transverse variable at $\tau_0$ moves radially outwards. There is no
entropy-flux across the flow lines so we get an idea of where the
initial multiplicity goes (once the transverse volume is properly
accounted for).  The slope of the flow line, $dr(t)/dt$, gives the
local radial flow velocity $v_r(r(t),t)$. Thus, the bending of the
flow lines seen in the QGP and HRG phases indicates acceleration of
the fluid. The straight flow lines observed in the MP  in turn
indicate a constant pressure and no acceleration.

The corresponding constant velocity contours at RHIC and LHC, which,
reading from left to right are $v_r=0.1, \ 0.2, \dots, 0.9$, are shown
in Figs.~\ref{fig:flowRHICb} and \ref{fig:flowLHCb}. 
The velocity contours illustrate how the
deflagration wave penetrates into the fireball as the radial pressure
gradient pushes the matter to collective motion. A kink at the boundary
of QGP and mixed phase (MP) results from disappearance of pressure
gradients in the MP. As there is no acceleration in the MP, the matter
continues to flow with radial velocity obtained already in QGP.
The reappearance of pressure gradients, when matter enters from MP to
hadron gas phase, is seen as a kink in the velocity contours at the
phase boundary.

The dense bundle of velocity contours starting around $r\gsim 10$ fm
results from the treatment of the nuclear surface. Even though we
take the densities smoothly to zero, the expansion is essentially to
vacuum and the velocities approach the velocity of light. This region
has no consequences for any physical observables, since it is outside
the decoupling surface in a region with vanishingly small densities.

Since the larger energy density leads to larger pressure gradients
at LHC than at RHIC, the transverse flow at LHC grows also much
stronger than at RHIC.  This is seen both in the larger radial
extent of the decoupling curve and the larger values reached in
transverse velocity at decoupling. We next discuss how this affects
qualitatively the transverse spectra of hadrons.

\subsubsection{Effects of transverse flow on hadron spectra}

It is well known that transverse flow broadens the $m_T$ spectra of
hadrons and, since heavier particles obtain a larger momentum
increase from the collective flow velocity, the effective
temperature $\Teff$, defined as the inverse of the $m_T$ slope of
the logarithm of a spectrum, and the average transverse momentum
\begin{equation}
\langle p_T^h \rangle \equiv
{{\int dp_T\,\,p_T\,dN^h/dydp_T}\over{\int dp_T\,\,dN^h/dydp_T}} \,,
\label{<pT>}
\end{equation}
increase with particle mass. Before discussing the actual results, we
demonstrate the effects of flow in a simplified system of homogeneous
cylinder with radius $R$, which decouples at $\tau=\tau_f=$~const.
In this case, if $\mu_h=0$, the Cooper-Frye formula (\ref{CooperFrye}) 
reduces to\footnote{
This form, Eq~(\ref{blastwave}), was used in a blast-wave fit
\cite{Schnedermann:1993ws}, where $v_r=(r/R_A)v_{{\rm max}}$ was
assumed and where the parameters $T$ and $v_{{\rm max}}$ were fitted
to reproduce the {\it shape} of the final hadron spectra.
}
\begin{equation}\label{blastwave}
\frac{dN}{dydm_T^2} = \frac{g}{2\pi} \tau_fm_T \sum_{n=1}^\infty
(\pm 1)^{n+1} \int_{0}^{R} dr \ r
I_0(n\gamma_rv_r\frac{p_T}{T}) K_1(n\gamma_r\frac{m_T}{T}),
\end{equation}
where the index $h$ has been suppressed. Integration over $r$ is trivial if
$v_r=$~const. In particular, if no transverse flow is present,
$v_r=0$, the equation (\ref{blastwave}) simplifies to
\begin{equation}\label{mtscale}
\frac{dN}{dydm_T^2} =
\frac{g}{4\pi} \tau_f R^2 m_T\sum_{n=1}^\infty (\pm 1)^{n+1}
   K_1(\frac{nm_T}{T})
\stackrel{m_T/T\gg1}{\longrightarrow}
\frac{g}{\sqrt{32\pi}}\tau_f R^2 \sqrt{m_T T} \, e^{-m_T/T} \,,
\label{mTscaling}
\end{equation}
where also the asymptotic behaviour at large $m_T/T$ is indicated.

\begin{figure}[htb]
\begin{minipage}[t]{70mm}
    \epsfysize 11.5cm \epsfbox{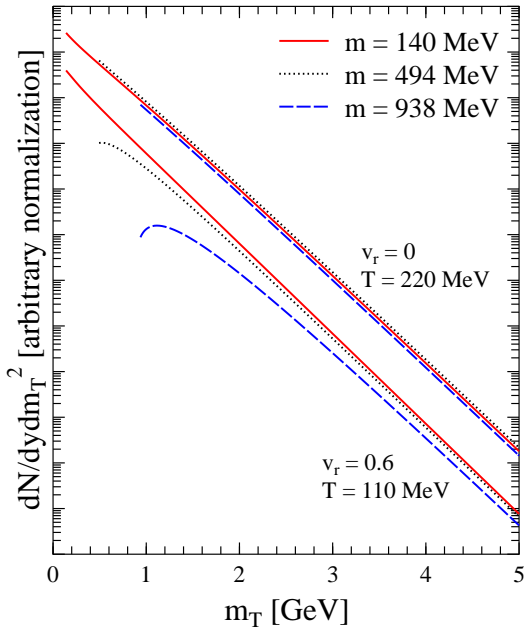}
\end{minipage}
 \hspace{20mm}
\begin{minipage}[t]{70mm}
 \vspace{-110mm}
 \begin{center}
  \begin{minipage}{65mm}
\caption{\protect\small
An example of how transverse flow affects the slopes of hadron spectra
in a simplified system which decouples at a constant proper time and
a constant radial flow velocity $v_r$. Two sets of spectra, computed
from Eq.~(\ref{blastwave}) but leaving the overall normalization arbitrary, are
shown for bosons of masses $m=$~140 (solid lines), 494 (dotted) and
938 MeV (dashed). The upper set of curves, where $v_r=0$ and
temperature $T=220$ MeV, shows $m_T$ scaling (for better visibility, 
the curves for 494 and 938 MeV are multiplied by 1.2 and 0.8, respectively).
In the lower set $v_r=0.6$ and $T=110$~MeV. At small $m_T$, the 
$m_T$ scaling is broken due to a non-zero $v_r$, but the spectral 
slopes are similar to the upper set at $m_T\gg T,m$.
}
  \label{fig:blast}
  \end{minipage}
 \end{center}
\end{minipage}
\hspace{\fill}
\vspace{-2.2cm}
\end{figure}

The upper set of curves in Fig.~\ref{fig:blast} is obtained using
Eq. (\ref{mtscale}), i.e. there is no transverse flow. In that case,
spectra of all particles, here three spin-0 bosons with masses
$m=140$, 494 and 938 MeV, are the same except that their $m_T$-spectra
start from particles mass (different curves are made visible by multiplying 
the $m=494$ and 938 MeV curves with 1.2 and 0.8, respectively).
This is known as the $m_T$-scaling. The overall normalization of the curves 
is left free, as we consider now the slopes only.

Non-zero transverse velocity breaks the $m_T$-scaling at $p_T\lsim
m$. This is illustrated by the lower set of curves in
Fig.~\ref{fig:blast} using constant transverse flow $v_r=0.6$. The
asymptotic behaviour of Eq.~(\ref{blastwave}) at $m_T/T\gg1$ shows
an exponential factor $\exp\{-m_T/T_{\rm eff}\}$ with $T_{\rm eff}
=\sqrt{\frac{1+v_r}{1-v_r}}T$. Thus the $m_T$ scaling of the slopes
is recovered asymptotically. In the example of Fig.~\ref{fig:blast} with
$v_r=0.6$, $T_{\rm eff}=2T$.  This explains why the slopes of the 
spectra are nearly the same in the cases $T=220$ MeV, $v_r=0$ and 
$T=110$ MeV, $v_r=0.6$.

This simple study very clearly demonstrates the role of the transverse
flow in the properties of transverse spectra of hadrons, and the
necessity to understand the dynamical evolution of the transverse
flow in detail for proper interpretation of measured slopes.

%%%%%%%%%%%%%%%%%%%%% FIGURE %%%%%%%%%%%%%%%%%%%%%%%%%%%%%%%%

%%%%%%%%%%%%%%%%%%%%% FIGURE %%%%%%%%%%%%%%%%%%%%%%%%%%%%%%%%

Let us turn to discuss the flow in more realistic hydrodynamical simulations,
presented in Figs. \ref{fig:flowRHICa}--\ref{fig:flowLHCb}.
There is an interesting interdependence between the EoS, spacetime
evolution of the flow and the slopes of thermal hadron spectra.  As
long as a phase transition is included in the EoS, the strength of
final flow on the decoupling surface, of fixed decoupling
temperature, is not very sensitive to the number of hadron 
resonance states included in the hadronic phase, while the actual 
spacetime evolution of the system may vary quite significantly.  
This can be understood as follows:
Reduction of the hadronic states in the EoS leads to a stronger
phase transition between the QGP and HRG, i.e. latent heat as well
as entropy ratio between the two phases increase. This makes the
deflagration from mixed phase to HRG stronger. It increases also
the pressure gradients in HRG since the reduction of hadronic states
leads to a larger sound velocity $c_s$ and the pressure is given by
$p = c_s^2 \varepsilon$. On the other hand, as discussed
in~\cite{ENRR}, the energy release from heavy resonance states
significantly slows down the decrease of temperature during the
evolution.  Thus, when we decrease the number of hadron resonance states in
HRG, it results in faster development of flow and drop in
temperature, leading to a shorter lifetime of the matter.  For larger
number of hadron resonance states the flow grows more slowly but the
energy release from heavy resonances leads to longer time for the
flow to build up.  In practice these effects compensate each other
and as a result, the slopes of the spectra are relatively
independent of the hadron content of the EoS.

There is, however, a relatively strong dependence of the effective
temperature, the inverse of spectral slope, on the decoupling
temperature~\cite{ENRR} once the hadron content of the EoS is fixed.
In Sec.~\ref{sec:results}, we shall illustrate the variation of the
hydrodynamical results in the range of decoupling temperatures
$T_{\rm dec}=150\dots 120$~MeV. In particular, with the given
pQCD+saturation initial state and with the relatively steep initial
transverse profile, we shall show that the RHIC spectra favour the
higher $T_{\rm dec}$ both at $\sqrt s=130$~GeV and 200 GeV.

\section{Hadrons from fragmentation of high-$p_T$ partons}\label{sec:frag}

\subsection{pQCD framework for computing high-$p_T$ spectra.}
  \label{sec:fragBaseline}

Inclusive hadron spectra in high energy hadronic or nuclear collisions
are computable in terms of collinearly factorized partonic hard parts
(cross sections in LO) $\hat \sigma_{ij\rightarrow f+k}$,
parton distribution functions (PDFs) $f_{i/A}, \, f_{j/B}$ and
fragmentation functions (FFs) $D_{f\rightarrow h}$, schematically as
\begin{equation}
 d\sigma^{AB\rightarrow h+X} = \sum_{ijk} f_{i/A}\otimes f_{j/B}\otimes
 \hat \sigma_{ij\rightarrow f+k}\otimes D_{f\rightarrow h},
\end{equation}
where the fragmentation function, $D_{f\rightarrow h}$, gives the
average multiplicity of $h$ hadrons produced from a parton $f$. The
NLO pQCD computation \cite{Borzumati:1995ib,Aurenche:1999nz}, which
requires NLO PDFs and FFs as well, provides a successful description
of the high-$p_T$ spectra in p+$\bar{\rm p}$ collisions at collider
energies. Also the leading-order considerations are useful and have
been shown to agree quite well with the data in the large-$p_T$
region, once the overall normalization, the $K$ factors, are extracted
from the data \cite{EH03}. We shall here make use of the latter: by
applying the energy-dependent $K$ factors of Ref.~\cite{EH03} and
including in the analysis also the recent p+p data from RHIC
\cite{Adams:2003kv}, we determine the $K$ factors at $\sqrt s=130$, 200 GeV
and also, through extrapolation, at the LHC energy $\sqrt s=5500$~GeV. Armed
with these and by including nuclear effects in the PDFs according to
the EKS98 parametrization \cite{EKS98}, we compute the baseline pQCD
spectra of high-$p_T$ hadrons for nearly central $A$+$A$ collisions at
RHIC and LHC. We do not make further modeling to account
for the Cronin effect, or any intrinsic $k_T$ effects in the PDFs or
in the FFs \cite{WANG,LEVAI,VITEV}.

As presented in detail e.g. in Ref. \cite{EH03}, inclusive spectrum
of the high-$p_T$ hadrons in $A$+$A$ collisions at an impact parameter
{\bf b} in the collinearly factorized LO pQCD formalism becomes
\begin{equation}\label{pQCDspectra}
\frac{dN^h_{AA}({\bf b})}{dm_T^2dy} = K(\sqrt s,Q,\mu_{\rm F}
)\cdot J(m_T,y)
\sum_f\int^1_{z_{\rm min}} \hspace{-0.cm}\frac{dz}{z^2}\,
D_{f\rightarrow h}(z,\mu_{\rm F}^2)
\frac{dN^f_{AA}({\bf b})}{dq_T^2 dy_f}
\end{equation}
where $f=g,u,d,...$ labels the different parton types and $h$ those of
hadrons. Transverse momentum and rapidity of the produced parton are
$q_T\ge q_{T0}$ and $y_f$, while $m_T$ and $y$ are the transverse
mass and rapidity of the hadron. 
The hadron of fractional energy $z=E_h/E_f$ is assumed to form collinearly
with its mother parton. Thus the kinematic variables are related as
\begin{eqnarray}
q_T & = & \frac{p_T}{z}J(m_T,y),\quad
         J(m_T,y) = (1-\frac{m^2}{m_T^2\cosh^2y})^{-1/2} \\
y_f & = & {\rm arsinh}(\frac {m_T}{p_T}\sinh y),
\quad{\rm and}\quad \frac{2m_T}{\sqrt s}\cosh y \le z
\le {\rm min}[1, \frac{p_T}{q_{T0}}J(m_T,y)]. \nonumber
\end{eqnarray}
We shall here apply the KKP fragmentation functions \cite{KKP} with a
scale choice $\mu_{\rm F}=p_T$. The inclusive spectra for the
production of parton type $f$ is obtained at LO from Eq.~(\ref{2parton}) by
integration \cite{EH03}, 

\begin{eqnarray}
\frac{dN^f_{AA}({\bf b})}{dq_T^2 dy_f}
&=& T_{AA}({\bf b})\frac{d\sigma^f_{AA}}{dq_T^2 dy_f}
= T_{AA}({\bf b})\int dy_1 dy_2 \sum_{\langle
kl\rangle}\frac{d\sigma}{dq_T^2dy_1dy_2}^{\hspace{-0.6cm}AB\rightarrow
kl+X} \cr
&&[\delta_{kf}\delta(y_f-y_1) + \delta_{lf}\delta(y_f-y_2)]
  \frac{1}{1+\delta_{kl}},
\label{parton_f}
\end{eqnarray}
where $T_{AA}$ is the nuclear overlap function from Eq.~(\ref{TAA}). We
define the nuclear PDFs by including the $x$- and $Q$-dependent EKS98
nuclear modifications \cite{EKS98} on top of the the CTEQ5
\cite{CTEQ5} PDFs of the free proton. Isospin effects in the nPDFs
are accounted for in the same manner as in Ref.~\cite{EKS98}. The
factorization/renormalization scale is again chosen as $Q=q_T$.

\begin{figure}[hbt]
\vspace{-2cm}
\begin{center}
   \includegraphics[height=12.0cm]{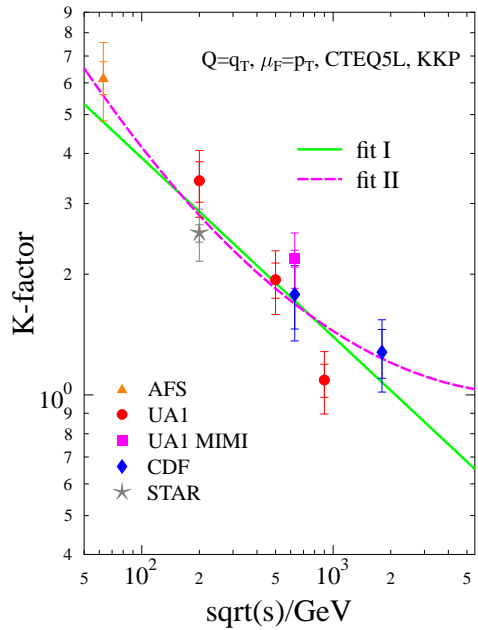}
\label{dndydmt}
\vspace{-1.0cm}
\begin{minipage}[h]{150mm}
\caption [1] {\protect\small
The $K$ factors obtained from fitting the computed leading-order
high-$p_T$ hadron spectra to the p+$\bar{\rm p}$ and p+p data. 
The fit I to $\ln K$, which is linear in $\ln \sqrt s$, see 
Eq.~(\ref{Kfits}), is shown by the solid line and the quadratic 
fit II by the dashed curve. The parameter values and error bar 
specifications can be found in the text. 
}
\label{fig:Kfactors}
\end{minipage}
\end{center}
\end{figure}

The factor $K(\sqrt s,Q,\mu_{\rm F})$ controlling the overall
normalization of the pQCD fragmentation spectrum in Eq.~(\ref{pQCDspectra}), is here
a phenomenological parameter which accounts for all higher-order
contributions in the partonic hard parts, PDFs and FFs (as well as for
any scale dependence arising from a truncation of the perturbation
series). This $K$ factor can be determined on the basis of the 
p+$\bar{\rm p}$ and p+p data. It should be emphasized that $K$ depends not only
on $\sqrt s$, but also the choices for the scales $Q$ and $\mu_{\rm
F}$, and on the PDF and FF sets used. The $K$ factors for
the parameters specified above have been obtained in
Ref. \cite{EH03} for various cms energies by using the data from the
AFS \cite{AFS}, UA1 \cite{UA1}, UA1 MIMI \cite{MIMI} and CDF
\cite{CDF} experiments. These results are shown in
Fig. \ref{fig:Kfactors}. As new input, we now include also the $K$ factor
corresponding to the recent STAR charged particle data from p+p collisions at $\sqrt
s=200$~GeV at RHIC \cite{Adams:2003kv}. The inner error bars shown
correspond to the statistical errors of our $K$-factor analysis
\cite{EH03}, for the outer error bars these have been added
in quadrature with the systematic errors of the data\footnote
{For CDF and UA1 MIMI we assume a 15 \% systematic error.}.
As discussed in detail in \cite{EH03}, the best agreement between the
shapes of the computed and measured spectra is obtained by
construction at $p_T\gsim 4\dots5$~GeV but the overall agreement is
surprisingly good over a wider $p_T$ range.

Based on the systematic decrease of $K(\sqrt s)$ in
Fig.~\ref{fig:Kfactors}, we extrapolate our LO pQCD reference spectra to
the LHC by making the following two simple parametrizations to the
obtained $K$-factors,
\begin{eqnarray}
{\rm I:}\quad \ln K&=& a + b\ln \sqrt s\\
{\rm II:}\quad \ln K    &=& a + b\ln \sqrt s + c(\ln \sqrt s)^2,
\label{Kfits}
\end{eqnarray}
where $\sqrt s$ is in GeV's. We get  $(a,b) = (3.4131; -0.44569)$ for
the fit I with $\chi^2=43.2$, and $(a,b,c) = (5.6206; -1.2139; \,
0.065615)$ with $\chi^2=40.6$ for the fit II. Both fits are also
shown in Fig.~\ref{fig:Kfactors}. Since systematic errors are not given
for all cases, only statistical errors (see \cite{EH03}) are
considered in these fits. At RHIC energies the $K$ factors are not
sensitive to the assumed form of parametrization, see
Fig.~\ref{fig:Kfactors}. For the pQCD spectra at the LHC, we shall
give an uncertainty band corresponding to the different $K$-factors
obtained from these fits. We note that a similar fit procedure was
introduced also in \cite{VITEV} but modeling in also a Gaussian
smearing from intrinsic-$k_T$ effects.

%%%%%%%%%%%%%%%%%%%%%% FIGURE %%%%%%%%%%%%%%%%%%%%%%%%%%%%%%%%
%%%%%%%%%%%%%%%%%%%%% FIGURE %%%%%%%%%%%%%%%%%%%%%%%%%%%%%%%%

\subsection{Energy losses for hard partons}\label{sec:Eloss}

To study the interrelation of the hydrodynamic and pQCD hadron spectra
at RHIC and at the LHC more meaningfully, we also consider effects
induced by the partonic energy losses to the high-$p_T$ pQCD fragmentation
spectra. First we note that the nuclear effects in the PDFs, which
are constrained to reproduce the nuclear DIS data through global DGLAP
fits, are not responsible for the factor $\sim 5$ suppression
discovered in the hadron spectra in central Au+Au collisions at RHIC
at $p_T\gsim 5$ GeV.  In fact, as shown in \cite{EH03}, the nuclear
modifications slightly (by 0...15\%) {\em enhance} the spectra below
$p_T\approx 13$~ GeV relative to p+p. The energy losses we consider
within the framework given in Refs. \cite{EHSW04,SW03}, by folding
quenching weights with parton production cross sections and FFs,
schematically written as
\begin{equation}
 d\sigma^{AA\rightarrow h+X} = \sum_{ijk} f_{i/A}\otimes f_{j/A}\otimes
 \hat \sigma_{ij\rightarrow f+k}\otimes P_f(\Delta E, L, \hat q)
 \otimes D_{f\rightarrow h}.
\end{equation}
The quenching weights, $P_f(\Delta E, L, \hat q)$, are generalized
probabilities for a hard parton $f$ to lose an amount $\Delta E$ of
its original energy through radiation induced by the medium over a
path length $L$. The transport coefficient $\hat q$ corresponds to
the average transverse momentum squared per unit pathlength gained by
the hard parton in traversing the matter. The transport coefficient
reflects the average number and energy densities of the system as
$\hat q\propto n\sim \epsilon^{3/4}$ \cite{Baier:2002tc}, and it
controls the energy loss.

We compute the pQCD spectra subject to energy losses according to the
formulation given in sec. 2.2 of Ref. \cite{EHSW04}. We do not repeat
the discussion here, but for illustrative purposes we note
that when the energy losses are taken into account, the analogue of
Eq.~(\ref{pQCDspectra}) can be written in terms of medium-modified
fragmentation  functions \cite{DMED} for central $A$+$A$ collisions as
\begin{equation}
\frac{dN^h_{AA}({\bf 0})}{dp_T^2dy} =
K\cdot J(m_T,y) \sum_f \int \frac{dz}{z^2}\, \int dr r \,d\phi
D_{f\rightarrow h}^{\rm (med)}(z,r,\phi,p_T^2)
\cdot [T_A(r)]^2
\frac{d\sigma ^f_{AA}}{dq_T^2 dy},
\label{ELOSSES}
\end{equation}
where the hard parton's original transverse momentum is $q_T =
J(m_T,y)p_T/z$, the distance of the hard parton's production point
from the center of the transverse plane is $r=|{\bf r}|$, and the
azimuthal angle between the hard parton's original transverse momentum
${\bf q_{T}}$ and the transverse vector ${\bf r}$ is $\phi$. The
medium-dependent fragmentation functions are defined as \cite{DMED}
\begin{equation}
D_{f\rightarrow h}^{\rm (med)}(z,r,\phi,p_T^2)
  = \int \frac{d\varepsilon}{1-\varepsilon}
    P_f(\varepsilon,L(r,\phi),\hat q) D_{f\rightarrow h}
    (\frac{z}{1-\epsilon},p_T^2),
\label{D_MEDeq}
\end{equation}
where $\varepsilon=\Delta E/E_i$ is the energy fraction lost by the
hard parton $f$ and the quenching weights are computed from those of
Ref. \cite{SW03}. The length $L$ traversed by the parton in the QGP now
depends on the transverse location of production and on the angle
$\phi$. In determining the path lengths $L$, a homogeneous transverse
disc profile of a radius $R_A$ is assumed \cite{EHSW04}.

In what follows, we compute the spectra with normalization fixed by
the $K$-factors determined above. As discussed in \cite{EHSW04}, the
obtained spectra are not sensitive to the (average) lifetime of the
plasma once it is of 4-5 fm/c or more, which is the case here, see
Figs.~\ref{fig:flowRHICa} and \ref{fig:flowLHCa}. In particular, in 5 \% 
most central Au+Au collisions at $\ssNN=200$ GeV we apply a 
large time-averaged transport coefficient $\hat q =\hat q_{200}= 10$~GeV$^2$/fm 
which was extracted in Ref. \cite{EHSW04} from the RHIC data in the region 
$p_T\gsim 5$~GeV. For the other energies and $A_{\rm eff}$ studied, we scale 
$\hat q$ according to 
\begin{equation}
\hat q(A,\ssNN)=\hat q_{200}(A/181)^{0.383}(\ssNN/200\,{\rm GeV})^{0.574}
\label{qhat_scale}
\end{equation}
on the basis of the average number
density scaling of our QGP initial conditions, as discussed in  \cite{EHSW04}.

The quenching weights applied here are computed in the eikonal
high-energy approximation where the lost energy $\Delta E$ can in
principle be arbitrarily large. Thus, with finite kinematics it may
happen that the probabilities are not always normalized properly to
unity within the parton energy range available,
\begin{equation}
\int_0^{q_T\cosh y} d(\Delta \, E) P_f(\Delta E, \hat q, L)<1.
\end{equation}
Although too large suppression can follow from this loss of
probability, this computation, however, gives us a lower limit of the
hadron spectra. An estimate of the upper limit can be obtained by
reweighting the quenching weights so that probability conservation is
always enforced, see \cite{Salgado:2003rv,EHSW04}. In our results for the
high-$p_T$ spectra (Figs.~\ref{fig:charged130}, \ref{fig:charged200},
\ref{fig:positive130}, \ref{fig:positive200}, \ref{fig:pi0200},
\ref{fig:chargedLHC}, \ref{fig:posLHC} ahead), we shall give an
uncertainty band corresponding to these two limits.

\section{Results}\label{sec:results}

We shall next compare the hydrodynamical and pQCD-based results with
the measured spectra in central and nearly central Au+Au
collisions at RHIC and make an extrapolation to Pb+Pb at the LHC. At
RHIC, the hydrodynamic spectra cover the small-$p_T$ region up to a
few GeV and  the pQCD spectra the region at $p_T\gsim4\dots5$~GeV. The
pQCD calculation of the hadron spectra is expected to become less
accurate at the lower end, below $p_T\lsim 4$~GeV (see \cite{EH03}),
and similarly the uncertainty in the tails of hydrodynamic results
grows with $p_T$. However, as will be seen below, the difference in
the slopes of the spectra from the hydrodynamic and the pQCD
calculations with energy losses in particular, is rather large. This
should reduce the uncertainty in the crossing of two contributions at
the LHC in particular, where the hydrodynamic spectra are shown to
dominate over a larger $p_T$ region than at RHIC.

\subsection{Comparison with RHIC data}

We first compare our results with the experimental data collected by
STAR \cite{Adams:2003kv, Adler:2002xw, Adams:2003xp, Adler:2002uv},
PHENIX \cite{Adler:2003au, Adcox:2001jp, Adcox:2001mf, Adcox:2002au,
Adler:2003cb, Adler:2003qi}, PHOBOS \cite{Back:2003qr,
unknown:2004zx} and BRAHMS \cite{Arsene:2003yk, Bearden:2004yx,
Bearden:2003hx} collaborations for the most central bins in Au+Au
collisions at $\ssNN=130$ and 200 GeV.
As explained in Sec.~\ref{sec:inicond}, the
centrality selection is accounted for by considering central $A_{\rm
eff}$+$A_{\rm eff}$ collisions.

\subsubsection{Transverse momentum spectra}

\begin{figure}[!ht]
\vspace{-1.7cm}
    \hspace{-1.5cm}
    \epsfxsize 100mm \epsfbox{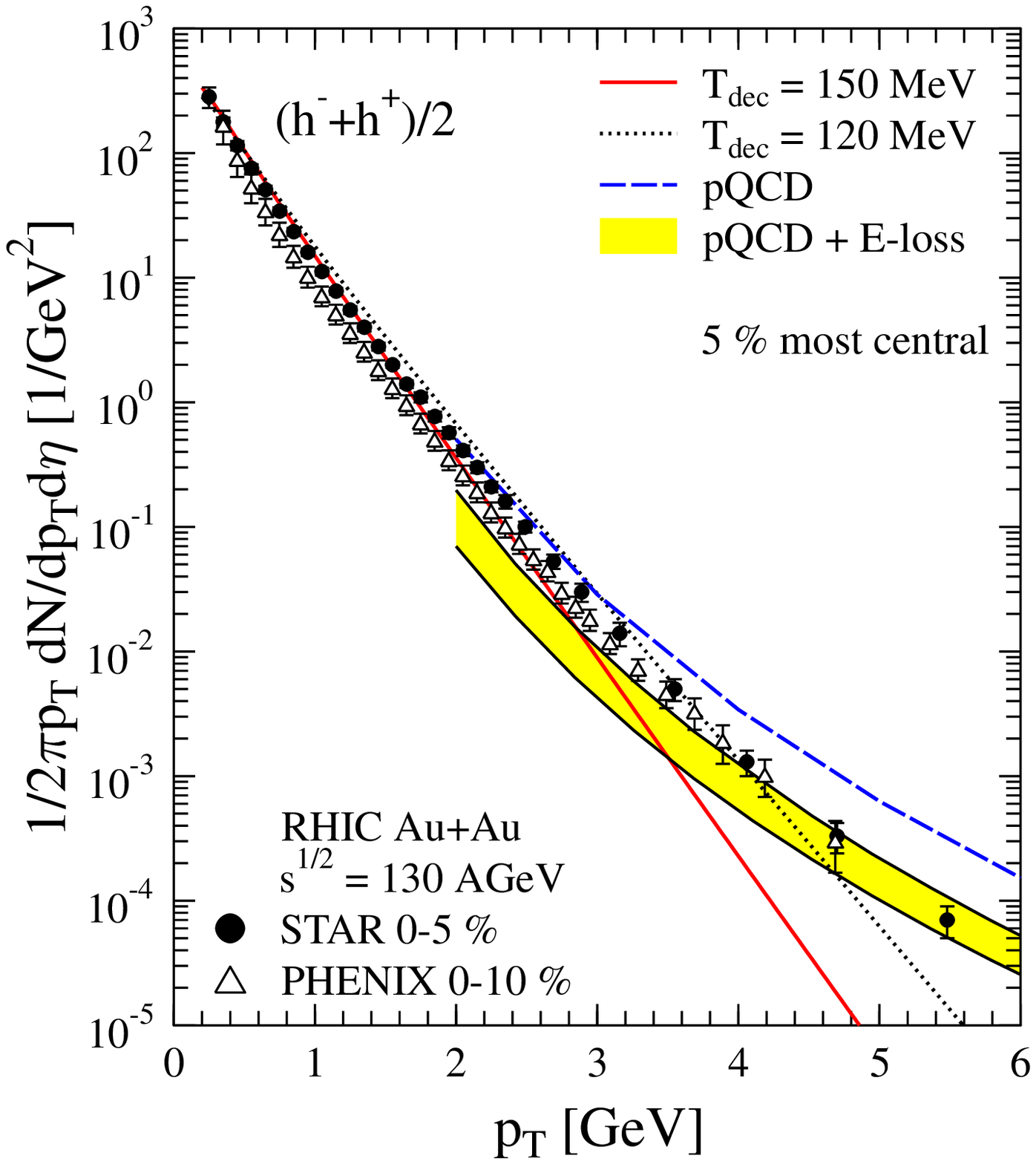} \hspace{-1.8cm}
	\epsfxsize 100mm \epsfbox{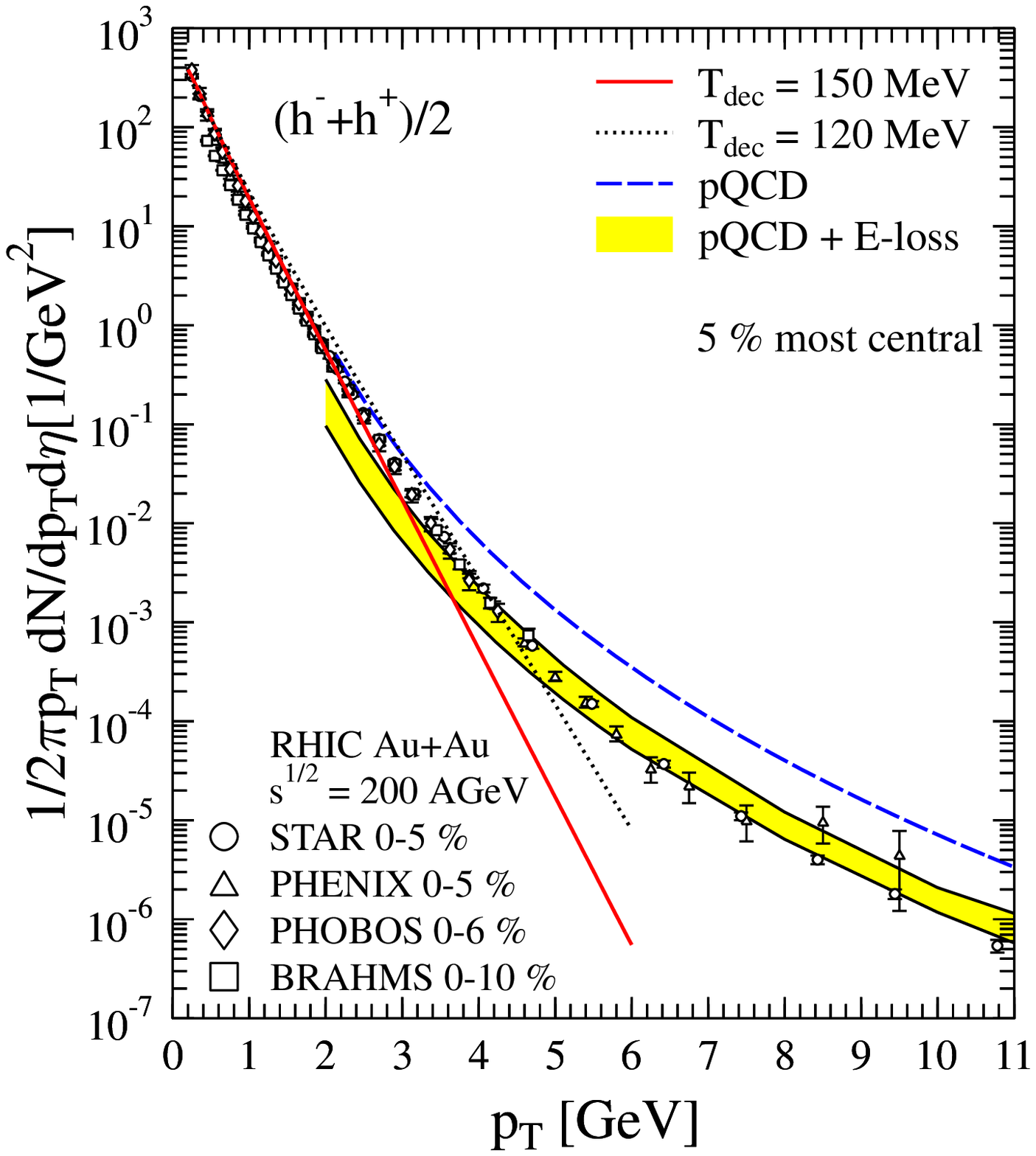}
 
 \vspace{-1.5cm}
\begin{minipage}[t]{75mm}
 \begin{center}
  \caption{\protect\small
Transverse momentum spectra of charged particles at $\eta=0$ (averaged
over $|\eta|\le0.1$) in 5 \% most central Au+Au collisions at
$\ssNN=130$~GeV. Our hydrodynamic results are shown for
$\Tdec=150$~MeV (solid line) and $\Tdec=120$~MeV (dotted line).  The
pQCD fragmentation results are shown with energy losses (shaded
band, see the text) and without (dashed line). The STAR data
\cite{Adler:2002xw} is plotted with the given total errors, and the
PHENIX data \cite{Adcox:2001jp} by adding the given statistical and
systematic errors in quadrature. Note the different centrality classes
of the two data sets.
}
  \label{fig:charged130}
 \end{center}
\end{minipage}
\hspace{\fill}
\begin{minipage}[t]{75mm}
 \begin{center}
  \caption{\protect\small
As Fig.~\ref{fig:charged130} but for 5 \% most central Au+Au collisions at
$\ssNN=200$~GeV. The data shown is taken by STAR \cite{Adams:2003kv}, PHENIX
\cite{Adler:2003au}, PHOBOS \cite{Back:2003qr} and BRAHMS
\cite{Arsene:2003yk} in the centrality classes indicated. We have
plotted the STAR and PHOBOS data with the given total error bars, the
PHENIX data by adding the given statistical and systematic errors in
quadrature and the BRAHMS data with the given statistical error bars.
}
  \label{fig:charged200}
 \end{center}
\end{minipage}
\end{figure}

First, we consider the $p_T$ spectra of charged particles measured
at central pseudorapidities $\eta$. Fig.~\ref{fig:charged130} shows 
the data taken in 5 \% most central Au+Au collisions at $\ssNN=130$
GeV by STAR \cite{Adler:2002xw} and PHENIX \cite{Adcox:2001jp}
collaborations. Similarly, Fig.~\ref{fig:charged200} shows the spectra
of charged particles at $\ssNN=200$ GeV measured by PHENIX
\cite{Adler:2003au}, STAR \cite{Adams:2003kv}, PHOBOS
\cite{Back:2003qr} and BRAHMS \cite{Arsene:2003yk} Collaborations.
The hydrodynamic results are computed accordingly \cite{ERRT}, using
two different decoupling temperatures, $\Tdec=150$ MeV (solid line)
and 120 MeV (dashed line). In addition to the strong and
electromagnetic decays, we have here included also the feed down from
the weak decays of hyperons, which affects the total number of charged
pions, kaons and (anti)protons on a few-per-cent level (see
Table~\ref{tab:GO} ahead). Note that our hydrodynamical results are
for 5 \% most central collisions while some data are taken in larger
centrality bins, as indicated in the figures.

%%%%%%%%%%%%%%%%%%%%% BEGIN FIGURE %%%%%%%%%%%%%%%%%%%%%%%%%%%%%%%%
%%%%%%%%%%%%%%%%%%%%% END FIGURE %%%%%%%%%%%%%%%%%%%%%%%%%%%%%%%%

Results in Figs. \ref{fig:charged130} and \ref{fig:charged200} show that
the normalization of the calculated hydrodynamical distributions is essentially
independent of the decoupling temperature and since the slopes
deviate only at larger transverse momenta, the total charged-particle
multiplicity is not very sensitive to the decoupling temperature
\cite{ERRT}.

If the system continues to undergo hydrodynamical evolution at
temperatures below 150 MeV, the collective flow will increase and
this is seen as an enhancement of high $p_T$ particles. For $\sqrt
s=200$ GeV, Fig.~\ref{fig:charged200}, the calculation with
$\Tdec=120$ MeV overshoots the experimental data already at $p_T\sim
1$ GeV, while a high decoupling temperature $\Tdec=150$ MeV describes
the results better. It is also clearly seen that the hydrodynamic
spectra cannot describe the RHIC data at larger transverse momenta,
$p_T\gsim2\dots3$ GeV. The same applicability region for the
hydrodynamical spectra at RHIC has been suggested also by the analyses
of azimuthal anisotropies, see e.g. \cite{Kolb:2003dz}.

Here it should be emphasized that we have not tuned the computation of
the initial state (e.g. the possible unknown constant in the saturation
criterion in Eq.~\ref{satcrit}) to fit the normalization to the data
but kept it as in the original predictions \cite{EKRT,ERRT}.
Given the robustness and simplicity of the pQCD+saturation model,
we obtain surprisingly good description for pion and kaon spectra. As
pions dominate the multiplicity, our results for integrated
observables, such as total multiplicity and transverse energy, are
also close to the experimental results. Spectra of protons and heavier
particles, especially at $\ssNN=200$ GeV, show details which are to be
addressed in more detail.

The dashed lines in Figs.~\ref{fig:charged130} and \ref{fig:charged200}
present the spectra of charged hadrons (charged pions, kaons and
(anti)protons included) from fragmentation of partonic jets as
predicted by Eq.~(\ref{pQCDspectra}) in the case of no energy losses.
It is seen that this prediction is clearly above the data when $p_T$
grows beyond 4-5 GeV. The solid curves bordering the shaded area
are the corresponding upper and lower limits of spectra computed
with energy losses as explained in Sec.~\ref{sec:Eloss}. In all
these pQCD based curves, we apply the $\sqrt s$-dependent $K$ factor
according to fit II in Eq.~(\ref{Kfits}). 

As observed earlier \cite{EH03} and in these figures, the high-$p_T$
spectrum is clearly suppressed relative to the baseline pQCD
fragmentation spectrum without energy losses. When the energy losses
are taken into account, the pQCD spectra agree with the data.
Apart from the $K$-factor, this agreement is obtained by
construction at $\ssNN=200$ GeV since the value of the transport
coefficient we use here was determined in \cite{EHSW04} by a fit to
the data at $p_T\gsim 5$~GeV. It is, however, quite interesting to see
how the pQCD fragmentation+energy loss spectrum in turn gradually
fails to describe the data at $p_T\lsim 4\dots5$~GeV. Thus, in the
intermediate-$p_T$ region, $2\dots3$ GeV $\lsim p_T\lsim 4\dots 5$ GeV
at RHIC neither the hydrodynamic nor pQCD description is sufficient to
describe the data alone but {\em both} are needed. We will return to
this interesting crossover region and the independency of the two approaches
in Sec.~\ref{hydvspqcd}.

%%%%%%%%%%%%%%%%%%%%% BEGIN FIGURE %%%%%%%%%%%%%%%%%%%%%%%%%%%%%%%%
\begin{figure}[!ht]
 \vspace{-0.7cm}
    \hspace{-1.7cm}
    \epsfxsize 100mm \epsfbox{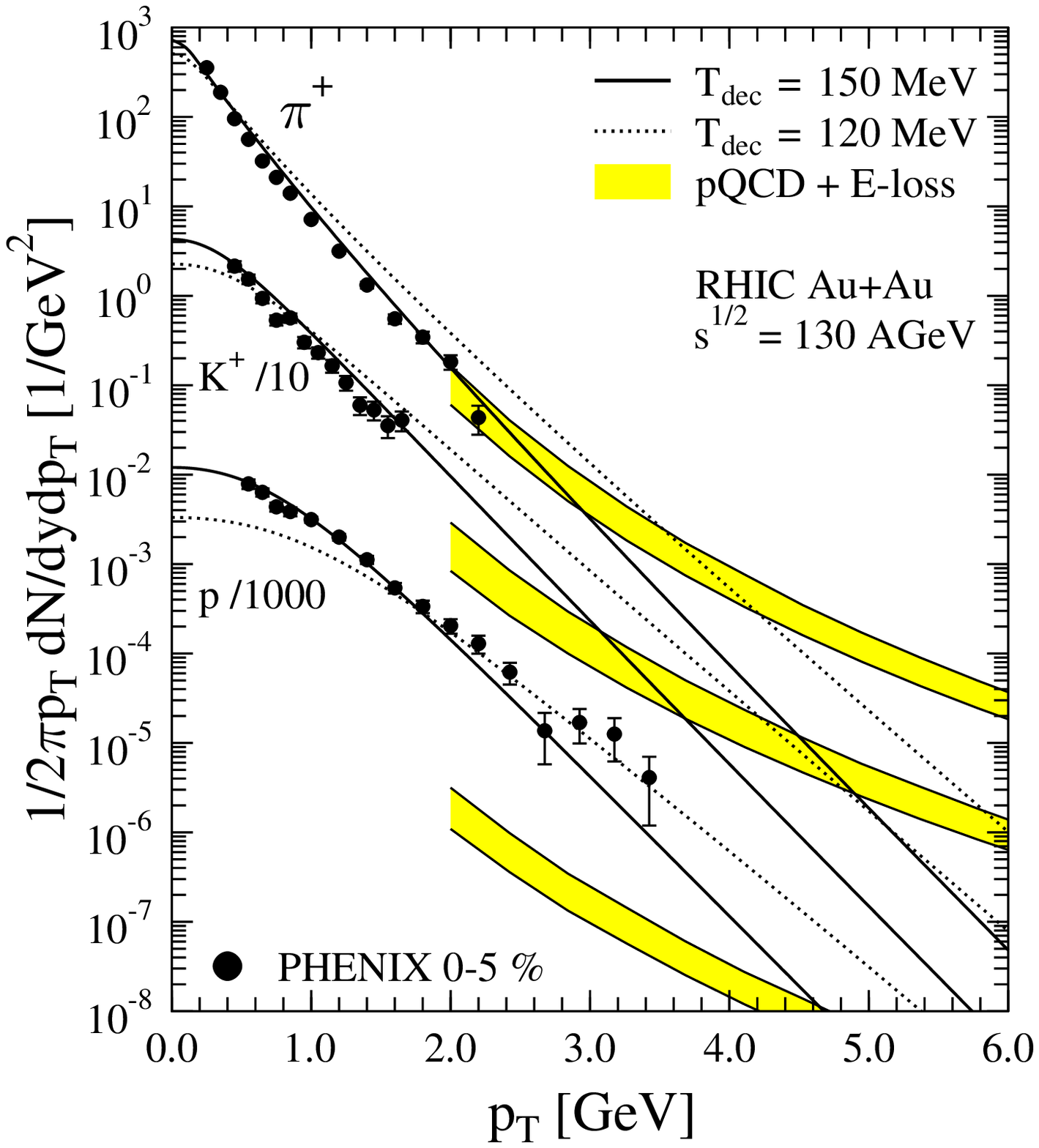}  \hspace{-1.8cm}
	\epsfxsize 100mm \epsfbox{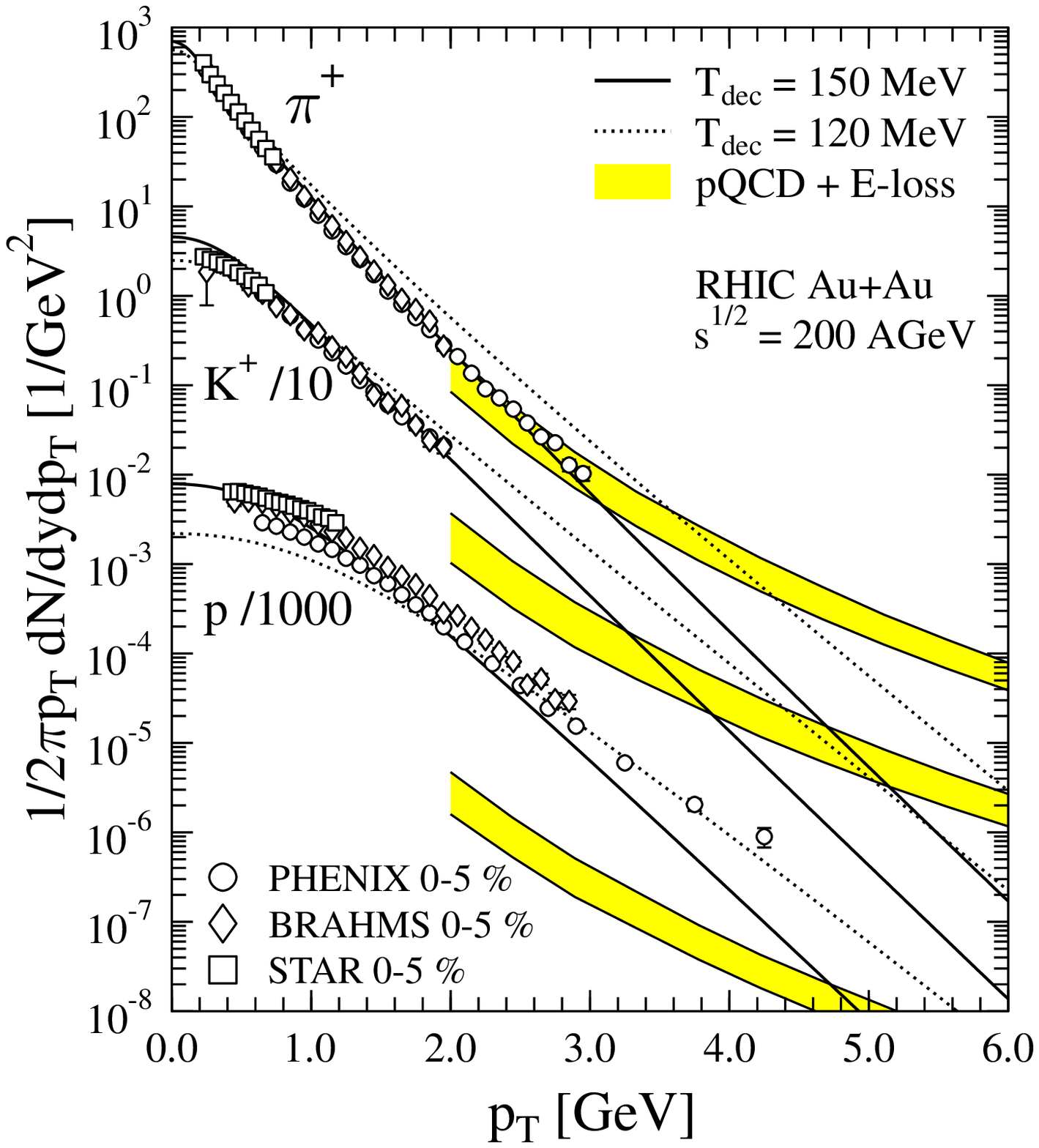}

 \vspace{-2.5cm}
\begin{minipage}[t]{75mm}
 \begin{center}
  \caption{\protect\small
Transverse momentum spectra of positive pions, positive kaons and
protons at $y=0$ in 5 \% most central Au+Au collisions at $\ssNN=130$~GeV.
The solid lines show our hydrodynamic results with $\Tdec=150$~MeV and
the dotted lines the results with $\Tdec=120$~MeV. The shaded bands
correspond to the pQCD fragmentation results with energy losses.
The PHENIX data \cite{Adcox:2001mf} is plotted with the given total
error bars. Note the scaling factors 10 and 1000 for kaons and
protons, respectively. Both the hydrodynamic result and the PHENIX
data contain the feed-down contributions from hyperons.
}
  \label{fig:positive130}
 \end{center}
\end{minipage}
\hspace{\fill}
\begin{minipage}[t]{75mm}
 \begin{center}
  \caption{\protect\small
As Fig.~\ref{fig:positive130} but at $\ssNN=200$~GeV. The PHENIX data
\cite{Adler:2003cb} and the BRAHMS data \cite{Bearden:2004yx,
Bearden:2003hx} are shown with statistical errors
and the STAR data \cite{Adams:2003xp} by the given total error bars. The 
hydrodynamic calculation and the PHENIX data are without the hyperon
feed-down contributions but the STAR and BRAHMS data contain the feed-down.
}
  \label{fig:positive200}
 \end{center}
\end{minipage}
\end{figure}
%%%%%%%%%%%%%%%%%%%%% END FIGURE %%%%%%%%%%%%%%%%%%%%%%%%%%%%%%%%

Next, we proceed to the $p_T$ spectra of identified hadrons at
midrapidities.  Fig.~\ref{fig:positive130} shows the
PHENIX data collected for positive pions, kaons and protons in 5\%
most central Au+Au collisions for $y=0$ at $\ssNN=130$~GeV
\cite{Adcox:2001mf}. Similarly, in Fig.~\ref{fig:positive200} STAR
\cite{Adams:2003xp}, PHENIX \cite{Adler:2003cb} and BRAHMS
\cite{Bearden:2004yx, Bearden:2003hx} data are shown at $\ssNN=200$~GeV.
Notice the scaling factors 10 and 1000 for kaons and protons,
respectively. The hydrodynamic spectra are computed with a {\em common}
decoupling temperature, i.e. the thermal and the kinetic freeze-out
take place simultaneously. The results with a high decoupling
temperature, $\Tdec=150$~MeV, are shown by the solid curves, and
those with a lower value, $\Tdec=120$~MeV, by the dashed ones. It
is seen that there is a substantial change in the normalization at
small $p_T$ for protons and kaons, indicating also a change in their
multiplicity $dN/dy$. For pions the change in the multiplicity with
$\Tdec$ is small but the slopes change considerably for all three
particles. Using a common high decoupling temperature $\Tdec = 150$
MeV (solid lines) our results describe both the normalization and the
slopes of the experimental spectra quite well whereas for
$\Tdec=120$ MeV (dotted lines) the normalization of protons and
kaons is too low and the $p_T$ dependence of all spectra is too
shallow.

The shaded bands in Figs.~\ref{fig:positive130} and \ref{fig:positive200} 
show the spectra of pions, kaons (divided by 10) and protons (divided by 1000)
calculated from pQCD+ fragmentation+energy loss. 
For the pion
spectra the hydrodynamic and the pQCD calculations complement nicely
each other:  the former describes the data at low and the latter at
high $p_T$'s. Even though tempting, it would be too naive to simply
add the two contributions without further considerations. 
E.g. one would expect the thermalization
assumption to fail in the large momentum tails of the thermal
distribution modifying the region where the dominance of thermal
part goes over to the dominance of the pQCD part. See the discussion in 
Sec.~\ref{hydvspqcd}. However, for protons
at $p_T\sim 3$~GeV, the pQCD fragmentation contribution is 
clearly smaller than the hydrodynamic one and also clearly 
below the measured proton spectra. We thus conclude that
pQCD fragmentation contribution as considered here
is not the origin of the intermediate-$p_T$ proton abundance.

The PHENIX data \cite{Adcox:2001mf} in Fig.~\ref{fig:positive130} include
the feed-down from weak decays of hyperons\footnote{
PHENIX has published a feed-down corrected spectrum of antiprotons in
minimum bias events \cite{Adcox:2002au}.}.
Our hydrodynamic spectra are here shown with the feed-down as well.
While for pions and kaons the hyperon feed-down is a small effect, for
protons it corresponds to a $\sim 30\%$ increase. On the other hand,
the hyperon feed-down is removed from the $\ssNN=200$ GeV PHENIX data
\cite{Adler:2003cb} in Fig.~\ref{fig:positive200}, but the STAR
\cite{Adams:2003xp} and BRAHMS \cite{Bearden:2004yx, Bearden:2003hx}
data include feed-down contributions. Our hydrodynamic results are
shown without feed-down.

%%%%%%%%%%%%%%%%%%%%% BEGIN FIGURE %%%%%%%%%%%%%%%%%%%%%%%%%%%%%%%%
\begin{figure}[hbt]
 \vspace{-1.8cm}
    \hspace{-1.5cm} 
    \epsfxsize 100mm \epsfbox{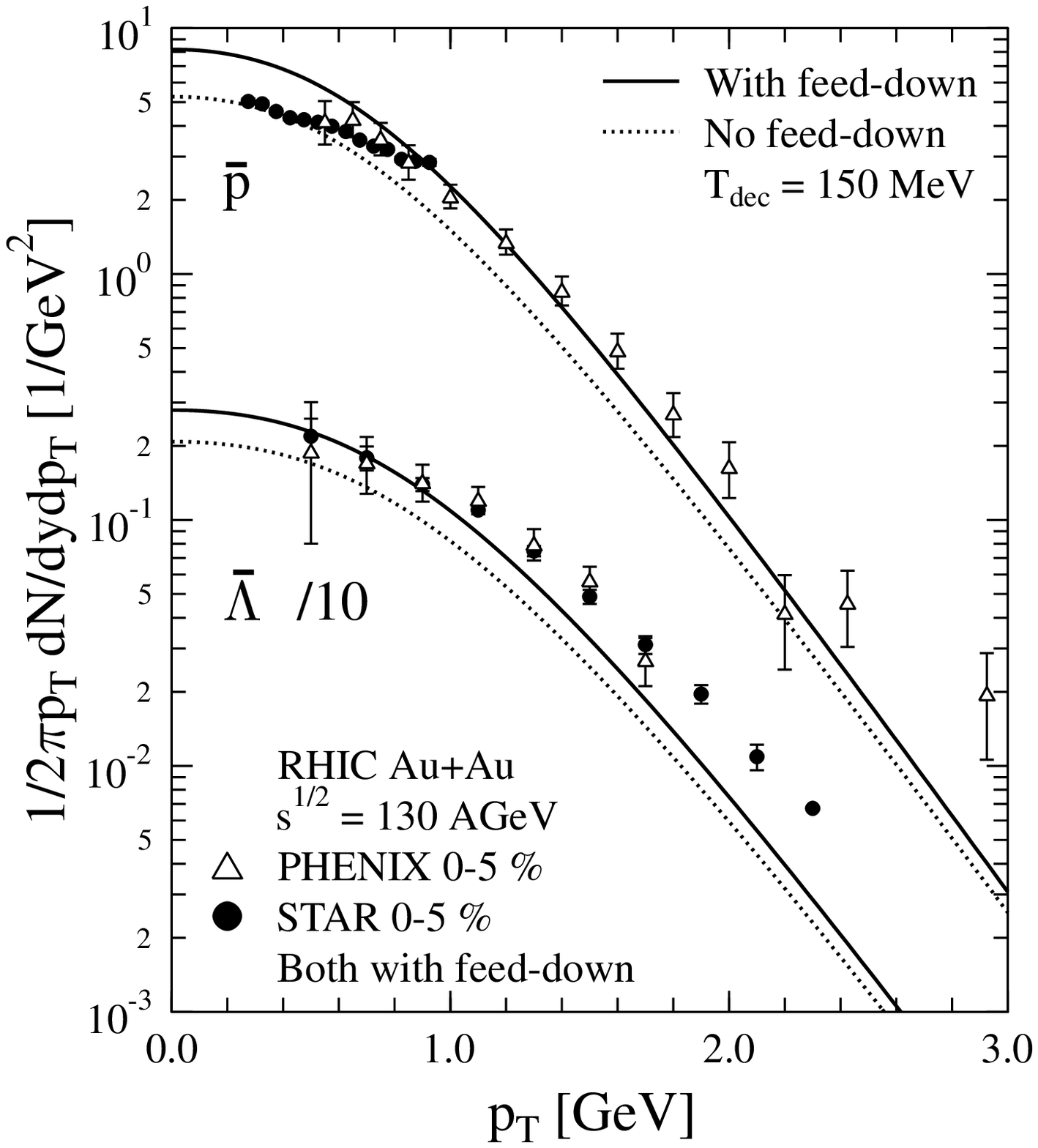}   
	
	\vspace{-11.1cm}\hspace{6.7cm}
	\epsfxsize 100mm \epsfbox{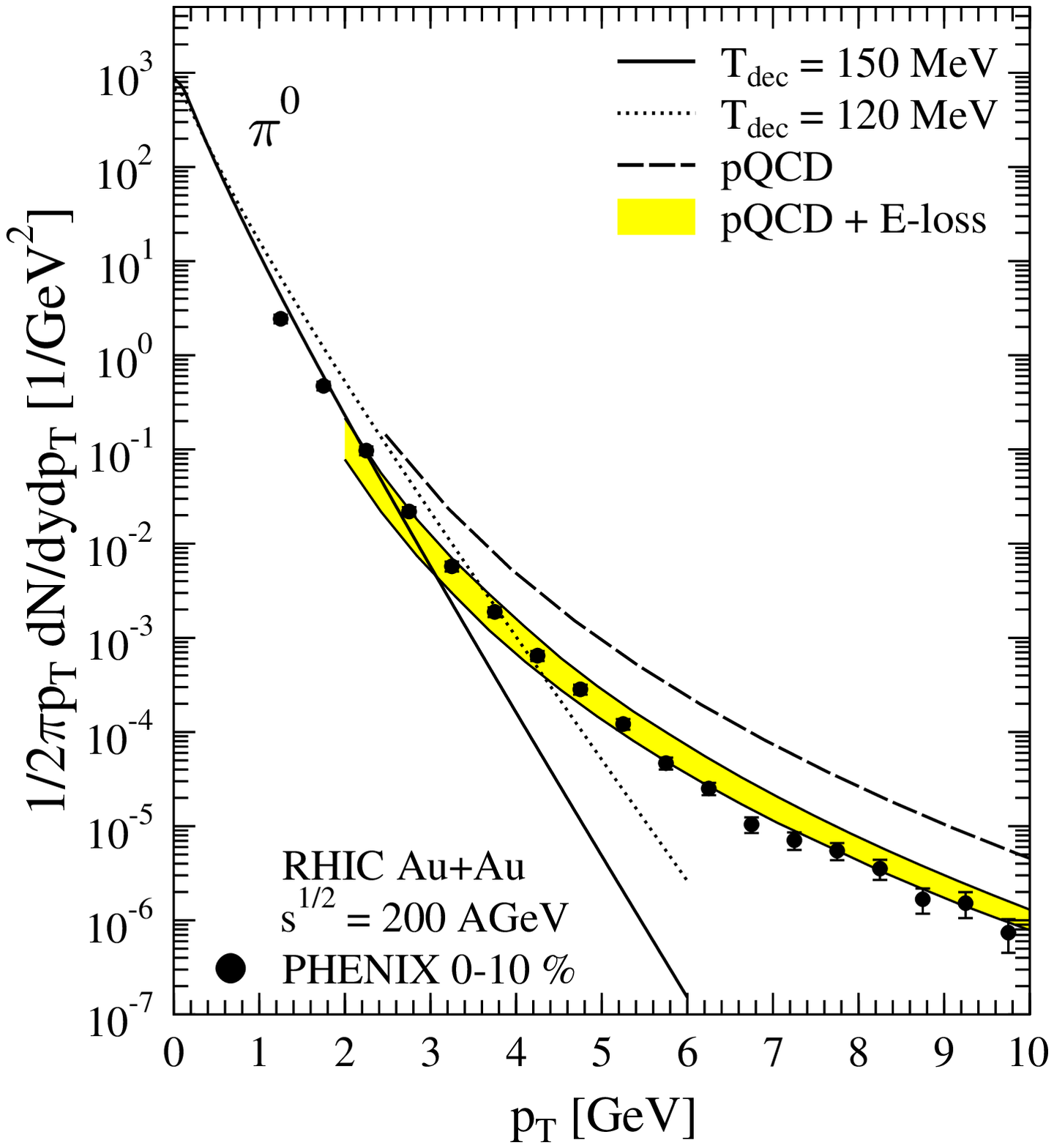}

    \vspace{-2.5cm}
\begin{minipage}[t]{75mm}
 \begin{center}
  \caption{\protect\small
Transverse momentum spectra of antiprotons and antilambdas
at $y=0$ in 5 \% most central Au+Au collisions at $\ssNN=130$~GeV.
Our hydrodynamic results are for $\Tdec=150$~MeV, the solid lines
showing them with hyperon feed-down contributions and the dotted lines
without. The PHENIX $\bar p$ \cite{Adcox:2001mf} and $\bar\Lambda$
\cite{Adcox:2002au} data and the STAR data \cite{Adler:2002uv} contain
the feed-down contributions. In the PHENIX $\bar p$ data the error
bars shown correspond to the given total errors, all others to the
statistical errors.
}
  \label{fig:lambda130}
 \end{center}
\end{minipage}
\hspace{\fill}
\begin{minipage}[t]{75mm}
 \begin{center} 
  \caption{\protect\small
Neutral pion production in 10 \% most central Au+Au collisions at 
$\ssNN=200$~GeV. Our hydrodynamic results are shown for
$\Tdec=150$~MeV (solid line) and $\Tdec=120$~MeV (dotted line).  The
pQCD fragmentation results are shown with energy losses (shaded
band, see the text) and without (dashed line). The data is taken 
by PHENIX \cite{Adler:2003qi} and shown with the given total error bars.
}
  \label{fig:pi0200}
 \end{center}
\end{minipage}
\end{figure}
%%%%%%%%%%%%%%%%%%%%% END FIGURE %%%%%%%%%%%%%%%%%%%%%%%%%%%%%%%%

Our hydrodynamic antiproton and antilambda spectra without (dashed
line) and with (solid line) the hyperon feed-down contributions are
shown in Fig.~\ref{fig:lambda130} for 5~\% most central Au+Au collisions
at $\ssNN=130$ GeV. The data in this figure are from STAR
\cite{Adler:2002uv} (circles) and PHENIX \cite{Adcox:2002au}
(triangles) in the corresponding centrality bin. Both data sets
include the hyperon feed-down contributions and should thus be
reproduced by the solid lines. Again, the agreement is
surprisingly good, except at the largest $p_T$ bins.

Since neutral pions produce the most important background for direct
and thermal photons \cite{Rasanen:2002qe,HPC_AURENCHE,Huovinen:2001wx}, 
an accurate description of neutral pion spectra
would be particularly important. In Fig.~\ref{fig:pi0200} we compare
our $\pi^0$ spectrum with that measured by PHENIX
\cite{Adler:2003qi} in 10 \% most central Au+Au collisions at
$\ssNN=200$ GeV. Again, the hydrodynamic computation is shown for
the low (dotted) and high (solid) $\Tdec$, and the pQCD
fragmentation+energy loss results, where we assume
$\pi^0=(\pi^++\pi^-)/2$, and the $K$ factor from the charged-particle 
p+$\bar{\rm p}$(p) data (Fig.~\ref{fig:Kfactors}),
are shown by the shaded band. The
conclusions here are the same as for charged pions: the high
$\Tdec$ describes the small-$p_T$ part well until $p_T\sim 3$~GeV,
while the pQCD+energy loss spectrum reproduces the high-$p_T$ part
$p_T\gsim 5$~GeV. In the intermediate $p_T$ region, $3\lsim
p_T\lsim 5$~GeV, both contributions are needed.

\begin{figure}[hbt]
 \begin{center}
   \vspace{-0.8cm}
    \epsfxsize 90mm \epsfbox{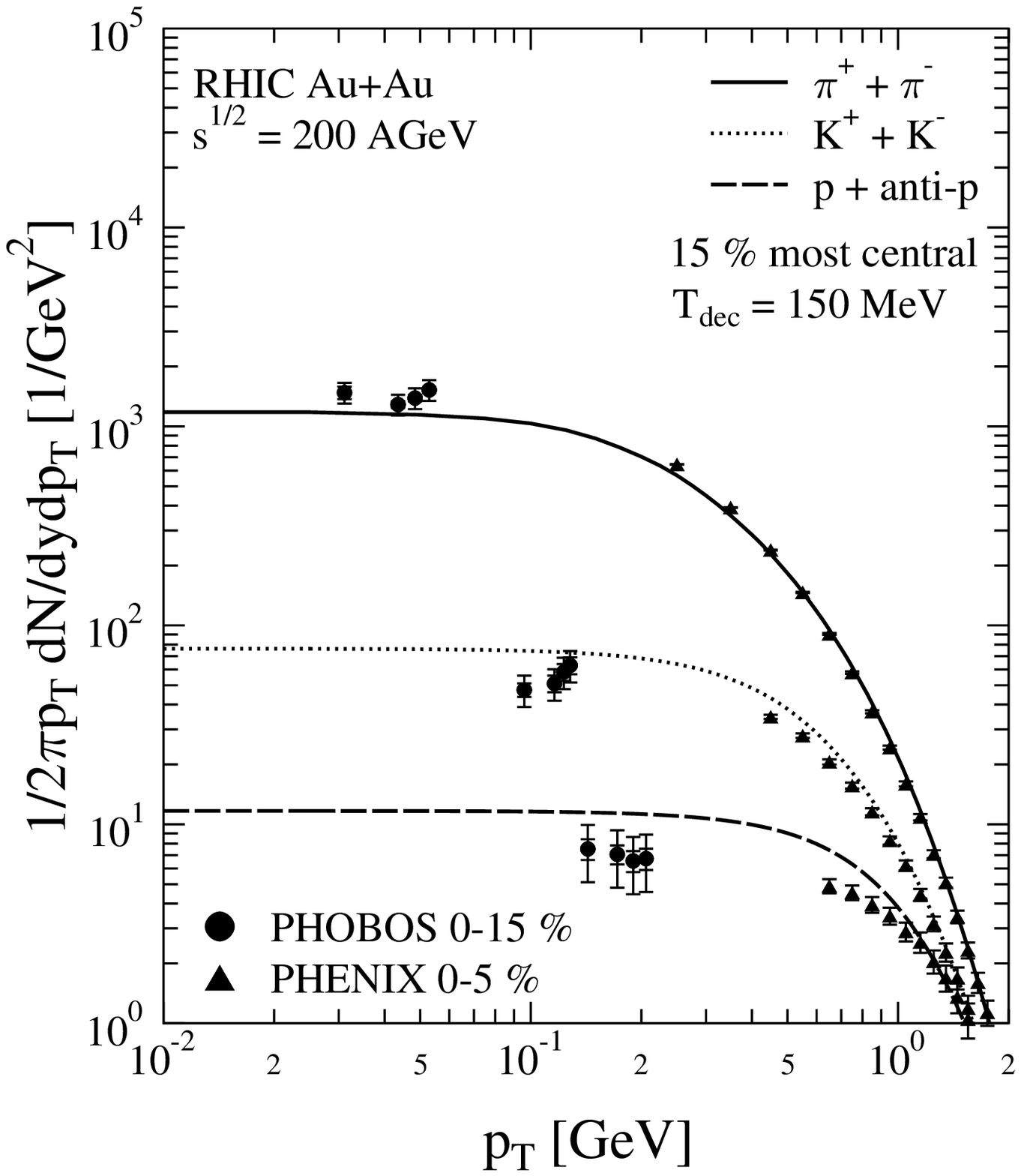}
   \vspace{-1.6cm}
  \caption{\protect\small
Transverse momentum spectra of charged pions (solid), charged kaons 
(dotted) and protons + antiprotons (dashed) at $y=0$ in 15 \% most
central Au+Au collisions at $\ssNN=200$~GeV in the ultralow-$p_T$
region measured by PHOBOS \cite{unknown:2004zx}. Our hydrodynamic
results are for $\Tdec=150$~MeV. The PHOBOS data is shown with
statistical and systematic error bars and the PHENIX data
\cite{Adler:2003cb} by adding the statistical errors in
quadrature. Note the different centrality classes in the two data sets.
}
  \label{fig:lowpt200}
 \end{center}
\end{figure}

Figure \ref{fig:lowpt200} shows experimental results for very low
transverse momentum spectra of hadrons measured by the PHOBOS
collaboration \cite{unknown:2004zx}.
%%%%%%%%%%%%%%%%%%%%% BEGIN FIGURE %%%%%%%%%%%%%%%%%%%%%%%%%%%%%%%%
%%%%%%%%%%%%%%%%%%%%% END FIGURE %%%%%%%%%%%%%%%%%%%%%%%%%%%%%%%%
Also the spectra measured by the PHENIX collaboration
\cite{Adler:2003cb} extending to larger $p_T$ are shown.  Both
collaborations have removed feed-down contributions from their data.
Our results for $\pi^++\pi^-$, K$^++$K$^-$ and p$+\bar{\rm p}$ are
plotted as solid, dotted and dashed line, respectively. 
The measured spectra are quite well reproduced.
On a log-scale the small-$p_T$
interval illuminates clearly that also spectrum of pions becomes flat
when $p_T<m_\pi$, which is also supported by the PHOBOS data. The
spectra of kaons and protons we obtain, are slightly above the
data. This will be seen in the multiplicities of these particles
discussed in the next section.

\subsubsection{Multiplicities and feed-down effects}

Experimental results for multiplicities, $dN/dy$, of $\pi^\pm$,
K$^\pm$, p and $\bar{\rm p}$ from PHENIX \cite{Adler:2003cb} and STAR
\cite{Adams:2003xp} collaborations with $\ssNN=200$ GeV are presented
in Table \ref{tab:expGO}. PHENIX has subtracted the feed-down from their
data, but STAR data include it. Pion and kaon multiplicities in these
experiments agree within the given errors. The difference between the
PHENIX and STAR results for (anti)protons, however, may require a more
detailed study of the feed-down effects.

%%%%%%%%%%%%%%%%%%%%% BEGIN TABLE %%%%%%%%%%%%%%%%%%%%%%%%%%%%%%%%
\begin{table}[h]
\begin{center}
\begin{tabular}{|c||c|c|c|c|c|c|}
\hline
$dN/dy$ & $\pi^+$ & $\pi^-$ & K$^+$ & K$^-$ & p & $\bar{\rm p}$ \\
\hline
PHENIX & $286.4\pm24.2$ & $281.8\pm22.8$ & $48.9\pm6.3$ &
         $45.7\pm5.2$  & $18.4\pm2.6$ & $13.5\pm1.8$ \\
STAR   & $322\pm32$ & $327\pm33$ & $51.3\pm7.7$ & $49.5\pm7.4$
       & $34.7\pm6.2$ & $26.7\pm4.0$ \\
\hline
\end{tabular}
\end{center}
\caption{\small Experimental results for $dN/dy$ in 5 \% most central Au+Au
collisions with $\ssNN=200$ GeV from PHENIX \cite{Adler:2003cb} and
STAR \cite{Adams:2003xp} Collaborations. The PHENIX results are corrected
for the feed-down, in the STAR results feed-down  is in the data.}
\label{tab:expGO}
\end{table}
%%%%%%%%%%%%%%%%%%%%% END TABLE %%%%%%%%%%%%%%%%%%%%%%%%%%%%%%%%
%%%%%%%%%%%%%%%%%%%%% BEGIN TABLE %%%%%%%%%%%%%%%%%%%%%%%%%%%%%%%%
\begin{table}[h]
\begin{center}
\begin{tabular}{|c||c|c|c|c|}
\hline
$\ssNN$ [GeV] & \multicolumn{2}{c|}{130} & \multicolumn{2}{c|}{200} \\
\hline
$\Tdec$ [MeV] & 120 & 150 & 120 & 150 \\
\hline
\hline
$dN/dy$(tot) & 1086 (1097) & 1141 (1178) & 1284 (1297) & 1352 (1396) \\
$dN/d\eta$(tot) & 960 (969) & 961 (988) & 1141 (1152) & 1148 (1181) \\
$dN/dy$(char) & 671 (680) & 687 (716) & 794 (804) & 815 (849) \\
$dN/d\eta$(char) & 597 (604) & 587 (608) & 710 (718) & 701 (726) \\
$dE_T/dy$(tot) [GeV] & 696 (696) & 709 (709) & 850 (850) & 868 (868) \\
$dE_T/d\eta$(tot) [GeV] & 625 (625) & 590 (592) & 769 (769) & 731 (733) \\
\hline
$dN^{\pi^+}/dy$ & 284 (287) & 269 (279) & 337 (340) & 319 (332) \\
$dN^{\pi^0}/dy$ & 312 (316) & 298 (312) & 370 (375) & 354 (370) \\
$dN^{\pi^-}/dy$ & 284 (289) & 269 (282) & 337 (342) & 319 (334) \\
$dN^{{\rm K}^+}/dy$ & 43.6 (43.7) & 53.0 (53.1) & 51.3 (51.3) & 62.4 (62.5) \\
$dN^{{\rm K}^-}/dy$ & 41.1 (41.1) & 50.3 (50.4) & 49.1 (49.1) & 60.1 (60.2) \\
$dN^{{\rm p}}/dy$ & 10.1 (14.0) & 20.5 (30.2) & 10.4 (14.5) & 23.0 (34.0) \\
$dN^{\bar{\rm p}}/dy$ & 3.7 (5.2) & 14.0 (21.1) & 5.0 (7.1) & 17.5 (26.4) \\
$dN^\Lambda/dy$ & 4.1 (5.1) & 9.9 (12.9) & 4.3 (5.4) & 11.2 (14.6) \\
$dN^{\bar\Lambda}/dy$ & 1.6 (2.0) & 7.2 (9.5) & 2.1 (2.7) & 9.0 (11.8) \\
\hline
$\bar{\rm p}/{\rm p}$ & 0.36 (0.37) & 0.68 (0.70) & 0.48 (0.50) & 0.76 (0.78)\\
$\bar\Lambda/\Lambda$ & 0.39 (0.39) & 0.73 (0.74) & 0.49 (0.51) & 0.80 (0.81)\\
\hline
$dN_{B-\bar B}/dy$  & \multicolumn{2}{c|}{16.6} & \multicolumn{2}{c|}{14.0} \\
\hline
\end{tabular}
\end{center}
\caption{\small Multiplicities and total transverse energies  without and with
feed-down contributions for 5 \% most central Au+Au collision with
$\ssNN=130$ and 200 GeV. Results with feed-down are presented in
parentheses. For details, see the text. Also the net-baryon number 
obtained from Eq.~(\protect\ref{eq:netbaryons}) is shown.}
\label{tab:GO}
\end{table}
%%%%%%%%%%%%%%%%%%%%% END TABLE %%%%%%%%%%%%%%%%%%%%%%%%%%%%%%%%

Table \ref{tab:GO} summarizes our hydrodynamic results for
integrated observables in 5\% most central Au+Au collisions at RHIC
energies $\sqrt s_{NN}=130$ and 200 GeV.  The results are quoted for
the decoupling temperatures $\Tdec=120$ and 150 MeV. The total and
charged particle multiplicities and the total transverse energies at
central rapidity are listed first.  Below these, we show the total
multiplicities for different hadron species.  The last three rows of
the table show our results for the antibaryon-to-baryon ratios for
protons and lambdas, and the total net-baryon number.

The first numbers correspond to results in which the hyperon feed-down
contributions have not been considered, i.e. we include all the
electromagnetic and strong decays  but regard those hadrons stable
which have only weak decay channels. In the results given in
parentheses the hyperon feed-down contributions  have been added. The
feed-down from $\Lambda$, in particular, has been included to  all the
other yields while the feed-down for $\Lambda$ comes from the weak
decays of $\Sigma$, $\Xi$ and $\Omega$. (Note that $\Sigma^0$ decays
electromagnetically.)

As seen in Table \ref{tab:GO}, the hyperon feed-down contributions
remain small for pions and kaons\footnote{
Kaon yields are not affected by the $\Lambda$ decays.}
but are quite significant for lambdas and especially for protons.
The number of thermal excitations of heavier hadrons grows with
increasing $\Tdec$ and some of the decay chains of these heavy hadrons
end to proton or lambda increasing their yields strongly. Table
\ref{tab:GO} shows that e.g. the number of $\bar\Lambda$'s goes up by
a factor of four when $\Tdec$ is raised from 120 MeV to 150
MeV. A further $\sim$30-40~\% increase to p and $\Lambda$ yields comes
from the hyperon feed-down while the ratios $\bar{\rm p}/{\rm p}$ and
$\bar\Lambda/\Lambda$ remain almost constant.

Our pion multiplicities in Table \ref{tab:GO}, computed with
$\Tdec=150$~MeV, agree with both STAR and PHENIX results in
Table~\ref{tab:expGO}. For this $\Tdec$ we get, however, a slight
excess of kaons over the STAR and PHENIX data, as can be seen also in
Fig.~\ref{fig:positive200}. For protons, our results with feed-down are
consistent with the STAR data but the corresponding results without
feed-down overshoot the PHENIX data.
Table \ref{tab:GO} shows clearly how sensitive the baryon
multiplicities are to the decoupling temperature, emphasizing the
necessity for a high $\Tdec$ here. 

The baryon multiplicities have also a slight dependence on the number 
of hadron resonance states included in the EoS. It is, however, not 
clear how many resonance states should be taken into account.
It depends e.g. on the nature and dynamics of the phase
transition and on the lifetime of the hadron gas. There is no
guarantee that the number ratios of hadrons at hadronization are the
thermal ones.  Since the (anti)quark and gluon structure of hadrons varies, 
formation of some states could be favoured over
others. In the hadron gas interactions among hadrons would drive the
densities towards the thermal ones. Since the lifetime of hadron gas
is not very long, and the temperature drops fast, especially the
formation of heavier states might not be very effective.

Even though the baryon multiplicities are very sensitive to
$\Tdec$, the net-baryon number remains independent of $\Tdec$, 
since we have incorporated the conservation of net-baryon
current in the hydrodynamic equations~(\ref{eq:hydro}). The
final net-baryon number thus equals the initial one shown in 
Table~\ref{tab:GO}. Also the net-proton number is almost independent of $\Tdec$,
the change is only a few per cent when $\Tdec$ changes from 150 to 120 MeV.
Thus, the measured net-proton number gives a further probe for the 
computed initial state, governed by the saturation scale in our framework.
As seen in Tables~\ref{tab:expGO} and \ref{tab:GO}, our results for the 
net-proton number agree with the STAR and PHENIX data very well. 

\subsection{From RHIC to LHC: spectra and integrated observables}

The benefit of a closed calculation for the initial state is that we
can predict the hadron spectra, multiplicities and transverse
energies in central $A$+$A$ collisions  also at the LHC. The initial
conditions are given in Table~\ref{tab:pQCD}. Without a dynamical
model for decoupling (see e.g. Refs.~\cite{Teaney:2001gc,Teaney:2001av})
, however, we cannot predict the value for the
decoupling temperature. Based on the RHIC results, we would expect 
$\Tdec$ to be in the 150 MeV range. To show the sensitivity of our 
LHC results to $\Tdec$, we shall quote them with $\Tdec=120$~MeV also.

%%%%%%%%%%%%%%%%%%%%% BEGIN FIGURE %%%%%%%%%%%%%%%%%%%%%%%%%%%%%%%%
\begin{figure}[tbh]
 \begin{center}
   \vspace{-1.4cm}
    \epsfxsize 100mm \epsfbox{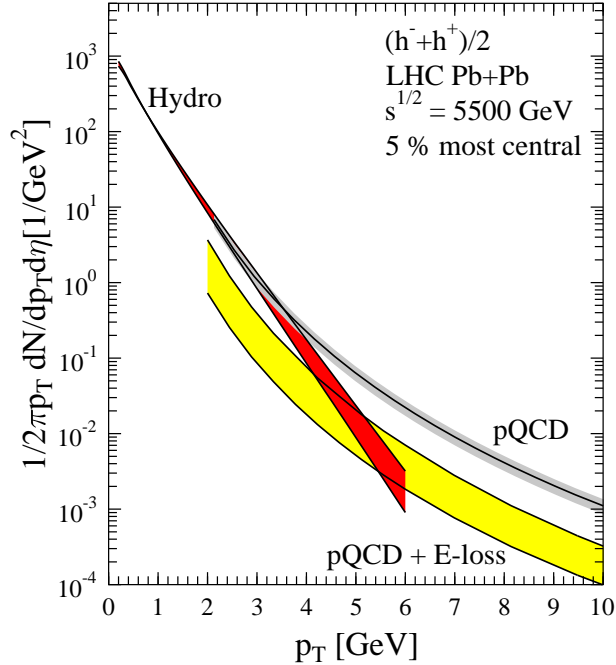}
   \vspace{-1.2cm}
    \caption{\protect\small
Our prediction for the transverse momentum spectra of charged hadrons
at $\eta =0$ in 5 \% most central Pb+Pb collisions at the LHC energy 
$\ssNN=5500$~GeV. The shaded band in the hydrodynamic results shows
the sensitivity of our results to $\Tdec$ in the interval $150\dots
120$~MeV. The solid curve labeled "pQCD" corresponds to the pQCD
fragmentation results where no energy losses have been taken into
account. The shaded band related to this is the estimated uncertainty
in the extrapolation of the $K$ factor to the LHC energy. The shaded
band labeled "pQCD + E-loss" describes the uncertainty in the pQCD
fragmentation results with energy losses, see the text for details.
}
  \label{fig:chargedLHC}
 \end{center}
\end{figure}
%%%%%%%%%%%%%%%%%%% end figure %%%%%%%%%%%%%%%%%%%%%%%%%%%

In Fig.~\ref{fig:chargedLHC} we show the charged particle spectra in 5 \%
most central Pb+Pb collisions at $\ssNN=5500$ GeV from our
hydrodynamical calculation with $\Tdec=150\dots 120$~MeV 
(the shaded band labeled "Hydro"), as well as from the
pQCD+fragmentation with and without energy losses. For the pQCD
spectra without the energy losses, the shaded uncertainty band around
the solid line stands for the difference between $K_1\approx0.65$ from
fit I and $K_2\approx1.03$ from fit II (see
Sec.~\ref{sec:fragBaseline}). In the pQCD calculation with the energy
losses included, we apply the latter $K$ factor (fit II has a slightly
better $\chi^2$ than fit I), and the large uncertainty band
corresponds to the difference between the reweighted and
non-reweighted results as discussed in the Sec.~\ref{sec:Eloss}.

We see that the hydrodynamical spectra are considerably flatter at
the LHC than at RHIC. This is due to the stronger flow developed at
the LHC. At the same time, the suppression in the pQCD spectra is
larger than at RHIC \cite{EHSW04}. The net result of these effects
is that the thermal part of the spectrum is expected to dominate
over the pQCD tail in a $p_T$ region up to $p_T\sim 4-5$ GeV that is
almost twice as wide as that at RHIC where the pQCD tails take over
already around $p_T\sim 2-3$ GeV. This will provide an opportunity
for more stringent tests of the thermal and hydrodynamical behaviour
of produced matter, including a wider $p_T$ range where the elliptic
flow should follow the hydrodynamical behaviour.

%%%%%%%%%% begin figure %%%%%%%%%%%%%
\begin{figure}[tbh]
 \begin{center}
   \vspace{-0.5cm}
    \epsfxsize 100mm \epsfbox{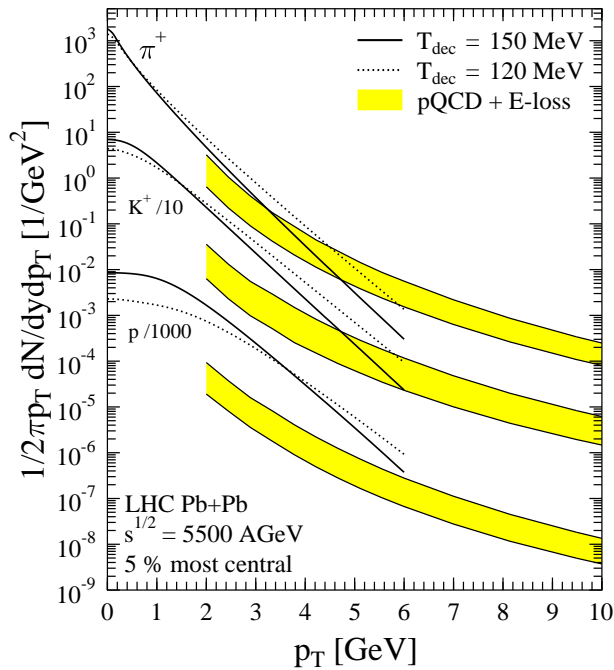}
   \vspace{-2.0cm}
    \caption{\protect\small
Transverse momentum spectra predicted for positive pions, positive kaons and
protons at $y=0$ in 5 \% most central Pb+Pb collisions at $\ssNN=5500$~GeV.
The solid lines show our hydrodynamic results with $\Tdec=150$~MeV and
the dotted lines the results with $\Tdec=120$~MeV. The shaded bands
correspond to the pQCD fragmentation results with energy losses.
}
  \label{fig:posLHC}
 \end{center}
\end{figure}
%%%%%%%%%%%%%%%%%%%%% END FIGURE %%%%%%%%%%%%%%%%%%%%%%%%%%%%%%%%

Figure \ref{fig:posLHC} shows our prediction for the spectra of
positive pions, kaons and protons at the LHC for the same
collision parameters as in the previous figure. The hydrodynamical
results are again given with the two different decoupling
temperatures, $\Tdec=150$~MeV (solid), and $\Tdec=120$~MeV (dotted).
We show the pQCD fragmentation results with energy losses using $K=1.03$.
The LHC results show similar systematics between the hydrodynamic 
and pQCD fragmentation results as at RHIC, Figs.~\ref{fig:positive130} 
and \ref{fig:positive200}; the crossover region
moves to larger $p_T$ for more massive particles. 
%%%%%%%%%%%%%%%%%%%%% BEGIN TABLE %%%%%%%%%%%%%%%%%%%%%%%%%%%%%%%%
\begin{table}[h]
\begin{center}
\begin{tabular}{|c||c|c|c|}
\hline
$\Tdec$ [MeV] & 120 & 150 & 160 \\
\hline
\hline
$dN/dy$(tot) & 4460 & 4730 & 4840 \\
$dN/d\eta$(tot) & 4120 & 4240 & 4290 \\
$dN/dy$(char) & 2760 & 2850 & 2900 \\
$dN/d\eta$(char) & 2560 & 2570 & 2600 \\
$dE_T/dy$(tot) [GeV] & 4010 & 4070 & 4110 \\
$dE_T/d\eta$(tot) [GeV] & 3790 & 3710 & 3680 \\
\hline
$dN^{\pi^+}/dy$ & 1170 & 1120 & 1120 \\
$dN^{\pi^0}/dy$ & 1290 & 1240 & 1240 \\
$dN^{\pi^-}/dy$ & 1170 & 1120 & 1120 \\
$dN^{{\rm K}^+}/dy$ & 175 & 214 & 218 \\
$dN^{{\rm K}^-}/dy$ & 175 & 214 & 218 \\
$dN^{{\rm p}}/dy$ & 25.8 & 70.8 & 88.1 \\
$dN^{\bar{\rm p}}/dy$ & 24.6 & 69.6 & 86.9 \\
\hline
$\bar{\rm p}/{\rm p}$ & 0.95 & 0.98 & 0.99 \\
\hline
$dN_{B-\bar B}/dy$  & \multicolumn{3}{c|}{3.11} \\
\hline\end{tabular}
\end{center}
\caption{Multiplicities and total transverse energies without
feed-down contributions for 5 \% most central Pb+Pb collisions at
$\ssNN=5500$ GeV. Also the net-baryon number 
predicted by Eq.~(\protect\ref{eq:netbaryons}) is given.
}
\label{tab:LHC}
\end{table}
%%%%%%%%%%%%%%%%%%%%% END TABLE %%%%%%%%%%%%%%%%%%%%%%%%%%%%%%%%

Integrated observables, total multiplicities and transverse
energies, and the multiplicities of different hadron species are
given in Table \ref{tab:LHC} for three different decoupling temperatures.
Even though the slopes of spectra at $p_T\gsim 1\dots 2$~GeV depend on
$\Tdec$, the total multiplicity and transverse energy are relatively
insensitive on $\Tdec$ \cite{ERRT}. Like at RHIC, only a $\sim 6$~\%
change is seen in $dN/dy$ and a few per cent change in $dE_T/dy$ when
changing $\Tdec$ from 150 to 120 MeV. Note that also at the LHC the
pion multiplicity remains quite insensitive to the decoupling temperature.
For kaons the change in multiplicity is $\sim 20$ \% (see also
Fig.~\ref{fig:posLHC} at small $p_T$) but for (anti)protons a factor
2.8.  For pions and kaons the change in multiplicity is negligible
if $\Tdec$ is raised from 150 to 160 MeV, whereas the (anti-)proton
multiplicity increases $\sim 25$ \%.

As seen in Tables \ref{tab:pQCD} and \ref{tab:LHC},
we predict an almost net-baryon free central rapidity region at
the LHC. The net-baryon number at $y=0$ is predicted 
to be down by a factor 5 relative to RHIC.
Although the proton and antiproton multiplicities are sensitive to
$\Tdec$, also the net-proton multiplicity, $dN^p/dy-dN^{\bar p}/dy$,
is practically independent of $\Tdec$ in the range $\Tdec=160\dots 120$~MeV. 
Due to the smallness of the net-proton number relative 
to the (anti)proton number, also the antiproton-to-proton ratio
$\bar p/p=0.98$ for $\Tdec=150$~MeV changes only by a few percent
when $\Tdec=160\dots 120$~MeV.

\subsection{How independent are hydrodynamic and pQCD spectra?}
   \label{hydvspqcd}

Interestingly, as seen in Figs.~\ref{fig:charged200} and 
\ref{fig:chargedLHC}, there is
quite a large difference between the slopes of the hydrodynamic and
pQCD spectra. Our results thus suggest that there is a rather narrow
$p_T$ interval in which  the dominant hadron production mechanism
changes from thermal emission at decoupling to independent partonic
jets suffering energy losses. It should be emphasized that we have
not attempted to construct an additive two-component model, where the
sum of the hydrodynamic and pQCD spectra would give the complete
spectra, valid over the whole $p_T$ range. First, in a
two-component model we should include only the thermalizing partons,
the region $\psat\le q_T\le q_{T{\rm th}}$, in the computation of the
hydrodynamic initial state. Correspondingly, only the
non-thermalizing partons, the region  $q_T\ge q_{T{\rm th}}$, should
be included in the pQCD fragmentation production  of hadrons. Second,
if we consider energy losses of jet-producing partons, we should add
the energy losses back to the hydrodynamic system. Since we do not
develop such a two-component model here, and since both computations
are now initiated with the same partonic cross sections, one should
ask to what extent our hydrodynamical and pQCD fragmentation spectra
in Figs. \ref{fig:charged200} and \ref{fig:chargedLHC} can be regarded mutually
independent, i.e. how much double counting there is if one simply
adds the two components together. In the regions where one of the
mechanisms dominates, this is obviously not an issue but it becomes
especially interesting in the crossover regions $3\lsim p_T\lsim
4$~GeV at RHIC and $5\lsim p_T\lsim 6$~GeV at the LHC, where both
components are of equal size. We shall argue below that in these
crossover regions (and of course at larger values of $p_T$), the
hydrodynamic and pQCD spectra are in fact biased to different
$q_T$-regions of originally produced partons and therefore the two
contributions now computed are to a good approximation additive
without serious double counting.

Consider first the RHIC spectra for 5\% most central Au+Au collisions
at $\ssNN=200$~GeV shown in Fig.~\ref{fig:charged200}. In the pQCD
fragmentation calculation without energy losses, hadrons at $p_T\sim
3$~GeV are dominantly produced by partons at $q_T\sim p_T/\langle
z\rangle\approx 1.7p_T\sim 5.1$~GeV, where $\langle z\rangle\approx
0.6$ follows from the slope of the partonic $q_T$ spectrum and the
shape of the (gluonic) fragmentation functions
\cite{EH03,Vogt:2004hf}. When energy losses are accounted for, the
dominating partonic momenta (for a fixed $p_T$) obviously grow
further, typically by, say\footnote{Unfortunately, our current
numerical energy loss set up does not allow for a more quantitative
estimate at $p_T\le q_{T0}$.},  a few GeV. Therefore the hadrons at
$p_T\sim 3$~GeV produced in pQCD fragmentation + energy loss can be
argued to be dominantly produced by partons originally at $q_T\sim
7$~GeV. In particular, the small-$q_T$ region $q_T\lsim 3.5$ does not
contribute much. On the other hand, we notice that in the
pQCD+saturation computation of the initial transverse energy, 95\% of
$\sigma\langle E_T \rangle$ (see Table~\ref{tab:pQCD}) comes from the
region $p_{\rm sat}\le q_T\le  3.6$~GeV. Were we to exclude the
region $q_T\ge q_{T{\rm th}}=3.6$~GeV from the calculation of the
initial conditions, the hydrodynamic spectra in
Fig.~\ref{fig:charged200}  would thus decrease only by a few per
cent. Similarly, exclusion of the region $q_T\le q_{T{\rm
th}}=3.6$~GeV from the pQCD fragmentation + energy loss calculation
should not visibly decrease the obtained pQCD spectra at $p_T\gsim
3$~GeV, or even at $p_T\gsim 2$~GeV. Thus we argue that to a good
approximation the computed hydrodynamic and  pQCD fragmentation +
energy loss spectra in Fig.~\ref{fig:charged200} at $p_T\gsim 2$~GeV,
and especially in the crossover region $3\lsim p_T\lsim 4$~GeV, can be
added together without significant double counting. Comparison with
the RHIC data also supports this conclusion.

At the LHC, in the 5~\% most central Pb+Pb collisions shown in
Fig. 15, hadrons at $p_T\sim 5$~GeV in the pQCD fragmentation
originate dominantly from $q_T\sim 8.5$~GeV partons when energy
losses are not accounted for. The average energy losses are now
larger than at RHIC, due to the denser system formed. Thus, we argue
that with the energy losses included, the $p_T=5$~GeV hadrons mainly
come from partons which are originally produced at, say, $q_T\sim
11\dots 12$~GeV. In particular, the region $q_T\lsim 6$~GeV should
not contribute much to the production of 5 GeV hadrons in the pQCD
fragmentation + energy loss computation. On the other hand, for the
LHC we notice that 90 \% of the initial $\sigma\langle E_T \rangle$
comes from $p_{\rm sat}\le q_T\le  6.6$~GeV and, correspondingly,
that exclusion of the region $q_T\gsim 6.6$~GeV from the initial
state computation would reduce the hydrodynamic spectra in Fig. 15 by
less than 10 \%. Thus, again in the crossover region $5\lsim p_T\lsim
6$~GeV and at higher $p_T$ at the LHC, the hydrodynamic and pQCD
fragmentation + energy loss spectra can be to a good approximation
regarded mutually independent, and the two components in Fig. 15
added together.

The basic assumption in the above discussion is that primary
small-$q_T$  partons ($q_T\le q_{T{\rm th}}$) thermalize and the
large-$q_T$ partons ($q_T\ge q_{T{\rm th}}$) go through the
thermalized matter just suffering energy loss but fragmenting
independently. We have argued above that if we set $q_{T{\rm th}}\sim
3.5$~GeV at RHIC and $q_{T{\rm th}}\sim 6$~GeV at the LHC, the
hydrodynamic and pQCD spectra do not significantly change (reduce)
from those shown in Figs.~\ref{fig:charged200} and 15. It should be
stressed, however, that the transition from thermalized to
jet-producing partons does not take place at any well-defined
$q_{T{\rm th}}$ but is gradual. There are partons in the region
around $q_{T{\rm th}}$ which are not fully thermalized but are not
well described by energy-losing independent high-energy partonic jets
either. Fig.~\ref{fig:charged200} suggests that our hydrodynamic and
jet fragmentation components are, when added together, not quite
enough to fully explain the RHIC data in the crossover region. This
leaves room for an extra contribution from the non-thermalized
partons near $q_{T{\rm th}}$ (e.g. hadronization with the thermal
ones) which the present treatment is not sensitive to.

What makes the final shape of the full hadron spectrum in the
crossover region very interesting is its dependence on the details of
thermalization. A precise experimental determination of the shape
provides an extra test for the understanding of thermalization of
primary partons.

\section{Discussion}\label{sec:conclusions}

%%%% YLEINEN SUMMARY
To summarize, we have computed the transverse momentum spectra of
hadrons in central and nearly central $A$+$A$ collisions at RHIC and
LHC in a broad $p_T$ range. For the low-$p_T$ spectra we apply the
hydrodynamic framework with pQCD+saturation initial conditions
\cite{EKRT,ERRT}. From the calculated primary production,
supplemented with the assumption of adiabatic expansion, we
correctly predicted the multiplicities at RHIC \cite{EKRT,ERRT,KJE_PVR_B}.  
For the high-$p_T$ part, we make a separate calculation folding together the
nuclear PDFs, partonic pQCD cross sections, energy losses and vacuum
fragmentation functions.

%%%% HYDROTULOKSISTA

Assuming thermalization at formation and no initial transverse flow,
we have a closed framework for computing the initial state for
radially expanding boost invariant hydrodynamics. Our results show
that the main features of the RHIC data on pion, kaon, (anti)proton
and (anti)lambda spectra, measured in (nearly) central Au+Au
collisions at $p_T\lsim 2\dots 3$~GeV, are quite well reproduced
with a single high decoupling temperature $\Tdec\simeq 150$~MeV.
The role of high $\Tdec$ for simultaneous chemical and kinetic
freeze-out is discussed. Also the feed-down contributions from
hyperons are studied in detail. We emphasize that the net-baryon
multiplicity is predicted by our calculation and our prediction
compares fairly well with the RHIC data. Unlike baryon or
antibaryon multiplicities, it does not depend on $\Tdec$, since the
net-baryon number is conserved in the hydrodynamical evolution.

The great benefit of a closed framework is that it can be easily
extended to the LHC energy. We have presented detailed predictions
for hadron spectra in (nearly) central Pb+Pb collisions at LHC which
depend both on the calculated initial conditions and the
hydrodynamic expansion. The overall multiplicity is less dependent
than the spectra on the details of expansion phase and the predicted
net-baryon number as a conserved quantity is completely fixed by the
calculation of primary production.

The overall behavior of our hydrodynamic results at RHIC looks
somewhat different from that at SPS energies. The experimental
results for Pb+Pb collisions at SPS are reproduced with a decoupling
temperature of the order $\Tdec \sim 120\ldots140$ MeV
\cite{Sollfrank_prc}, the lower values somewhat favoured by the $p-\bar
p$ spectrum. That there is a difference in the behaviour at SPS and
RHIC may not be so surprising since the systems start at quite
different densities. At RHIC the matter stays in the plasma state
longer and as the matter enters the hadron phase it flows faster
than at SPS.

One interesting detail, not studied in this work, is how the
transverse profile of the initial energy density \cite{PH_QM02}
affects the flow and the determination of $\Tdec$. As explained in
Sec.~\ref{sec:inicond}, here we consider a binary collision profile
with a fixed $p_{\rm sat}(A,\ssNN)$ and fixed $\tau_0=1/p_{\rm sat}$.
In a central $A$+$A$ collision, the saturation scales can, however, be
expected to grow towards the denser region and decrease towards the
edges \cite{EKT}. This would lower the density at center and
increase it at the edges. Also, the center of the system should
form earlier than the regions around it. This may have an effect on
the transverse profile of initial energy density, as the build-up of
transverse collective motion can start earlier at center
\cite{KHHET}.  Furthermore, the form of the initial transverse
profile should in fact depend also on the cms-energy due to parton
production dynamics in local saturation \cite{EKT}.  These studies
are left for later work.

It is quite interesting that with one common decoupling temperature,
one can reproduce the gross features of the low--$p_T$ RHIC data so
well. Certain details, in particular the fall of the proton
spectrum from hydrodynamic calculation below the data at $p_T\sim
3$~GeV need, however, improvements. As can be estimated in
Figs.~\ref{fig:positive130} and \ref{fig:positive200}, lowering of
$\Tdec$ to 140 MeV would improve the proton spectrum but it would somewhat
deteriorate the agreement of the pion and kaon spectra with data. This
points to the need of separate chemical and kinetic decoupling in
describing the details of all spectra simultaneously. Studies with
separate chemical and kinetic decoupling where the stable particle
numbers are fixed after chemical freeze-out, show indeed that the
spectra of pions and kaons become almost independent of the kinetic
decoupling temperature $T_{\rm dec,kin}$, whereas the (anti)proton
spectra widen with decreasing $T_{\rm dec,kin}$
\cite{HIRANO_TSUDA,Kolb:2002ve}.

%%%% pQCD FRAGMENTATION+E-LOSS-LASKUSTA
In our collinearly factorized LO pQCD calculation of the high-$p_T$
hadron spectra with absolute normalization, we make use of the
$K$ factor's $\sqrt s$ systematics demonstrated in \cite{EH03}. On top
of this baseline calculation, we also add nuclear effects in the PDFs
\cite{EKS98} and the energy losses of hard partons in the framework of
eikonalized quenching weights \cite{SW03,EHSW04}.
The amount of energy losses is controlled by the
time(path)-averaged transport coefficient  $\hat q$. The value of
$\hat q$ for central Au+Au collisions at $\ssNN=200$~GeV  was
determined in \cite{EHSW04} by fitting the observed {\em relative}
suppression in the region where hadron-type dependence disappears.
The current work, supplementing \cite{EHSW04}, thus provides an
important  cross-check of the overall normalization of the computed
spectra.

Our pQCD results, extended down to $p_T\sim 2\dots4$~GeV at RHIC,
show very interesting relation to the hydrodynamic results.  As seen
in Figs.~\ref{fig:charged130}--\ref{fig:positive200}, the pQCD
fragmentation + energy loss spectra of charged hadrons and pions
merge with the hydrodynamic one quite smoothly in the $p_T$ region
where the hydrodynamic calculation starts to fall below the data.
As the slopes in these two spectra are quite different, the region
of crossing from one mechanism to the other is not sensitive to
moderate changes in either contribution. As discussed in
Sec.~\ref{hydvspqcd}, the two results can be to a good approximation
considered independent in the crossover region and at higher $p_T$.

In particular, we can confirm that the difference between the proton
data at $p_T\sim 3$~GeV and the hydrodynamic results for $\Tdec=150$~MeV
cannot be explained by pQCD fragmentation + energy loss results considered 
here. As discussed above, a
separate chemical and kinetic freeze-out seems to keep the spectra
of kaons and pions almost independent of but widens the proton
spectrum so that the present data which ends at $p_T\approx 3.7$~GeV
can be described with hydro calculation. It would be extremely
interesting if the measured $p_T$ range of the proton spectrum could
be extended up to 6\ldots7~GeV. We would expect the pQCD
calculation of jet fragmentation with energy loss to reproduce the
data in that region and the possible deficit of predicted spectra,
if any, should show up in the $p_T$ interval from 4 to 6~GeV.

As discussed in detail above, our prediction for the high-$p_T$
spectra at the LHC is subject to uncertainties arising from the
extrapolation of the $K$ factor and transport coefficient into a new
energy and density regime, and from the normalization of the
quenching weights.  In particular, the uncertainties related to
using the quenching weights with finite kinematics are quite large
up to $p_T\sim 100$~GeV \cite{EHSW04}. The uncertainty of the $K$
factor can be reduced by scale-optimized NLO calculations
\cite{Aurenche:1999nz} and by applying the latest NLO nPDFs
\cite{deFlorian:2003qf} and, ultimately, by comparison with the
hadron spectra to be measured in p+p and p+$A$ collisions at the
LHC. Implementation of quenching weights into the NLO pQCD
framework will be a challenging task for the future. The most
urgent need would obviously be the inclusion of finite kinematics in
the computation of the quenching weights. Also the effects of 
detailed hydrodynamic evolution with transverse
expansion of the dense partonic system should be implemented into
the energy loss calculation.
\\ \\
{\bf Acknowledgements}
\\ \\
We thank P. Huovinen, J. Rak, D. Rischke, C. Salgado, K. Tuominen and U. Wiedemann
for discussions. This work was financially supported by the Academy of 
Finland, projects no. 50338 and 206024. H. Honkanen gratefully acknowledges 
the financial support from the Wihuri foundation.

\end{document}